\documentclass[%
preprint,
superscriptaddress,
preprintnumbers,
nobibnotes,
nofootinbib,
 amsmath,amssymb, 
 aps, prd,
]{revtex4-1}

\usepackage{cancel}
\usepackage{accents}
\usepackage{mciteplus,slashed}
\usepackage{amssymb,cancel,amsmath}
\usepackage{dcolumn}
\usepackage{bm}
\usepackage{soul}
\usepackage[caption=false]{subfig}
\usepackage{appendix}
\usepackage{physics}
\usepackage{booktabs}
\usepackage{comment}
\usepackage[T1]{fontenc}	
\usepackage{csvsimple}
\usepackage[bookmarksnumbered=true]{hyperref} 
\hypersetup{
    colorlinks = true,
    linkcolor = blue,
    anchorcolor = blue,
    citecolor = blue,
    filecolor = blue,
    urlcolor = blue
    }
\usepackage[section]{placeins}
\usepackage[capitalise]{cleveref}
\usepackage{booktabs}
\usepackage{graphicx}
\usepackage{mathrsfs}
\usepackage{relsize}
\graphicspath{{Figures/}}
\AtBeginDocument{
\heavyrulewidth=.08em
\lightrulewidth=.05em
\cmidrulewidth=.03em
\belowrulesep=.65ex
\belowbottomsep=0pt
\aboverulesep=.4ex
\abovetopsep=0pt
\cmidrulesep=\doublerulesep
\cmidrulekern=.5em
\defaultaddspace=.5em
}
\usepackage[dvipsnames]{xcolor}
\usepackage[normalem]{ulem}
\usepackage{fontawesome} 
\usepackage[force]{feynmp-auto}
\usepackage{bbm}
\usepackage{MnSymbol}
\usepackage{comment}
\usepackage[force]{feynmp-auto}

\def\iu{\mathrm{i}}
\def\e{\mathrm{e}}
\DeclareSymbolFont{cmbrightop}{OT1}{cmbr}{m}{n}
\DeclareMathSymbol{\sfPsi}{\mathalpha}{cmbrightop}{9}

\definecolor{wildstrawberry}{rgb}{1.0, 0.26, 0.64}

\begin{document}

\title{Manifest Gauge Invariance for Structure Dependent \\
Radiative Corrections to Processes Involving  Atoms and  Nuclei}

\author{Ryan Plestid}
\author{Mark B.\ Wise}
\affiliation{Walter Burke Institute for Theoretical Physics, California Institute of Technology, Pasadena, CA 91125, USA}

\date{\today}

\preprint{CALT-TH/2025-012}

\begin{abstract}
  Radiative corrections to reactions involving atoms or nuclei can become sensitive to the structure of the bound state. Generically, one encounters correlation functions of multiple currents which must satisfy Ward identities. At intermediate steps, however the Ward identities are obscured, and often violated by physically motivated approximation schemes. In this paper we outline a method to construct a representation of the aforementioned correlators that manifests gauge invariance in the limit of a heavy target (i.e., when recoil energy can be neglected). This representation then enables manifestly gauge invariant approximation schemes. Furthermore, the proposed representation naturally separates the largest contributions that dominate scattering amplitudes in the limit of a heavy constituent (e.g., proton) mass. We analyze elastic electron scattering from nuclei in detail, and also discuss radiative corrections to processes mediated by the weak interaction. 
\end{abstract}

\maketitle 

\section{Introduction \label{Intro} }
Electromagnetic corrections to reactions involving atoms and nuclei are phenomenologically important both because {\it i)} in some instances they can be large, and {\it ii)} their control is often a limiting factor  for precision applications (e.g., for fundamental physics) \cite{Ent:2001hm,Hill:2010yb,Tomalak:2020zfh,Rocco:2020jlx,Lovato:2020kba,Tomalak:2021hec,Tomalak:2022xup,Seng:2022cnq,Ruso:2022qes,Hill:2023acw,Hill:2023bfh,Cirigliano:2023fnz,Cirigliano:2024nfi,Cirigliano:2024rfk,Borah:2024ghn,VanderGriend:2025mdc,Afanasev:2023gev,Richardson:2023vyf,Seng:2024zuc,Combes:2024pvm}. It has long been appreciated that ``box diagrams'', those involving excited intermediate states of the bound-object, present theoretical difficulties due to their inherent dependence on the structure and spectrum of the nuclear or atomic target \cite{BOTTINO1966192,Bethe:1971es,Lin:1972ba,Friar:1974bn,Offermann:1991ft,Afanasev:2005mp,Gorchtein:2011mz,Arrington:2011dn,Seng:2018qru,Pasquini:2018wbl,JeffersonLabHallA:2018cmf,Kutz:2022hka,Jakubassa-Amundsen:2023wuz,Jakubassa-Amundsen:2024dai,Gennari:2024sbn}. 

Although the resultant amplitudes must satisfy Ward identities (related to gauge invariance), at intermediate steps of the calculation these identities may be obscured. This is far from an academic point, since one must often resort to approximations (either due to hadronic modeling of currents and the Hamiltonian or numerical evaluation of nuclear/atomic matrix elements), and without manifest gauge invariance the approximant to the amplitude may then violate the Ward identities. 

In what follows we outline a procedure to obtain a manifestly gauge invariant representation of correlators involving electromagnetic currents. The method relies on the neglect of target recoil energy, but does not require the constituent masses to be heavy. We find that the full correlator can be split apart into multiple separately gauge invariant pieces. By manipulating the formally gauge invariant amplitude, using identities that follow from current conservation, we arrive at a representation that is guaranteed to satisfy the Ward identities {\it even after} currents, the Hamiltonian, and nuclear matrix elements are approximated. We apply our results to a few relevant phenomenological examples including elastic electron scattering, and the weak interaction mediated process of radiative muon capture. 

\section{Elastic electron scattering}
Consider the elastic  scattering of electrons from a nuclear target,
\begin{equation}
    e(\vb{p}) A(\vb{0})  \rightarrow e(\vb{p}') A(\vb{Q})~. 
\end{equation}
In what follows we will neglect target recoil (taking $\vb{Q}^2\ll M_A^2$, with $M_A$ the target mass, such that the energy $\omega=Q_0=0$), but {\it will not} expand in the nucleon mass $M$.  For simplicity we will work in the lab frame and suppress spin labels, 
The matrix element has an order-by-order expansion in the electromagentic coupling constant $e^2$, 
\begin{equation}
    \mathcal{M}= e^2 \mathcal{M}^{(0)}  + e^4 \mathcal{M}^{(1)} + \ldots~,
\end{equation}
with the leading-order piece being given by
\begin{equation}
    \label{M_LO}
    \begin{split}
        \mathcal{M}^{(0)} &=  \bar{u}(p') \gamma^\mu u(p) \frac{1}{\vb{Q}^2}  \mel{A(\vb{Q)}}{J_\mu}{A(\vb{0})}\\
        &= \bar{u}(p') \gamma^\mu u(p) \int \frac{\dd^4 q}{(2\pi)^4} (2\pi)^4\delta^{(4)}(q-Q) \frac{-1}{q^2} \langle J_\mu(\vb{q}) \rangle ~,
     \end{split}
\end{equation}
where $Q_\mu=(0,\vb{Q})$ has no energy transfer, $q_\mu = (\omega,\vb{q})$, and $J_\mu = (\rho,\vb{J})$. The electron spinors are $u(p)$ for the initial electron and $\bar{u}(p')$ for the final electron. We use a convention such that the operator $J_\mu(\vb{q})$ injects momentum $\vb{q}$ into the system, with
\begin{equation}
    J_\mu(\vb{q}) = \int \dd^3x ~\e^{\iu \vb{q}\cdot \vb{x}} J_\mu(\vb{x},t=0)~. 
\end{equation}
This differs from typical conventions for Fourier transforms. 
The notation involving angle brackets is defined as, 
\begin{equation}\label{langle-rangle-def}
    \langle J(\vb{q}_1)\ldots J(\vb{q}_n) \rangle \equiv  \frac{\mel{A(\vb{p}')}{J(\vb{q}_1)\ldots J(\vb{q}_n)}{A(\vb{p})} }{\braket{A(\vb{p}')}{A(\vb{p}+\sum_i \vb{q}_i)}}~,  
\end{equation}
such that $\langle J(\vb{q}_1)\ldots J(\vb{q}_n) \rangle$ is a smooth function of $\vb{q}$. In what follows we use a non-relativistic normalization $\braket{A(\vb{p}')}{A(\vb{p})}=(2\pi)^3 \delta^{(3)}(\vb{p}'-\vb{p})$. For fixed $\vb{p}$, the right-hand side of \cref{langle-rangle-def} is independent of $\vb{p}'$. In the limit of infinite target mass the right-hand side is also independent of $\vb{p}$, because the velocity of the nucleus remains zero, hence \cref{langle-rangle-def} is a function of only the variables $\vb{q}_1 \ldots \vb{q}_n$. 

At second order in perturbation theory we encounter two different time orderings of the currents. This leads to (using Feynman gauge photon propagators), 
\begin{equation}
    \label{M_NLO}
    \begin{split}
    \mathcal{M}^{(1)} =  \int \frac{\dd^4 q_1}{(2\pi)^4}\frac{\dd^4 q_2}{(2\pi)^4}
    (2\pi)^4\delta^{(4)}&(q_1+q_2-Q) \\
    &\times \qty[\bar{u}(p') \gamma^\nu  \qty(\frac{1}{\slashed{p}-\slashed{q}_1 -m_e}) \gamma^\mu u(p)] \frac{1}{q_1^2} \frac{1}{q_2^2} H_{\mu\nu}(q_1,q_2)~,
    \end{split}
\end{equation}
where $q_{1\mu} = (\omega_1,\vb{q}_1)$ and $q_{2\mu} = (\omega_2,\vb{q}_2)$.   
\vfill
\pagebreak 
Pictorially we have 
\bigskip
\begin{equation*}
    \mathcal{M}^{(1)}~=~~ 
     \raisebox{-15pt}{
    \begin{fmffile}{matrix-el-box}
    \begin{fmfgraph*}(16,8)
    \fmfcurved
    \fmfleft{i1,i2}
    \fmfright{o1,o2}
    \fmf{double}{i1,v1,v5,v2,o1}
    \fmf{plain}{i2,v3,v6,v4,o2}
    \fmf{photon,tension=0,label=$q_1$,l.side=left}{v1,v3}
    \fmf{photon,tension=0,label=$q_2$,l.side=right}{v2,v4}
    \fmfv{label=$A(\vb{0})$}{i1}
    \fmfv{label=$A(\vb{Q})$}{o1}
    \fmfv{label=$e(\vb{p})$}{i2}
    \fmfv{label=$e(\vb{p}')$}{o2}
\end{fmfgraph*}
\end{fmffile} } 
    \quad~~
    + 
    \quad~~
     \raisebox{-15pt}{
    \begin{fmffile}{matrix-el-box-crossed}
    \begin{fmfgraph*}(16,8)
    \fmfcurved
    \fmfleft{i1,i2}
    \fmfright{o1,o2}
    \fmf{double}{i1,v1,v5,v2,o1}
    \fmf{plain}{i2,v3,v6,v4,o2}
    \fmffreeze
    \fmf{photon,tension=0}{v1,v4}
    \fmf{photon,tension=0}{v2,v3}
    \fmfv{label=$A(\vb{0})$}{i1}
    \fmfv{label=$A(\vb{Q})$}{o1}
    \fmfv{label=$e(\vb{p})$}{i2}
    \fmfv{label=$e(\vb{p}')$}{o2}
\end{fmfgraph*}
\end{fmffile} }
    \quad~~
    + 
    \quad~~
     \raisebox{-15pt}{
    \begin{fmffile}{matrix-el-box-seagull}
    \begin{fmfgraph*}(16,8)
    \fmfcurved
    \fmfleft{i1,i2}
    \fmfright{o1,o2}
    \fmf{double}{i1,v1,o1}
    \fmf{plain}{i2,v3,v5,v4,o2}
    \fmffreeze
    \fmf{photon,tension=0}{v1,v4}
    \fmf{photon,tension=0}{v1,v3}
    \fmfv{label=$A(\vb{0})$}{i1}
    \fmfv{label=$A(\vb{Q})$}{o1}
    \fmfv{label=$e(\vb{p})$}{i2}
    \fmfv{label=$e(\vb{p}')$}{o2}
\end{fmfgraph*}
\end{fmffile} }
\quad\quad,\\[25pt]
\end{equation*}
where $q_1$ and $q_2$ flow into the nucleus, and the last diagram is the seagull vertex discussed below.

The Hamiltonian is defined such that $H\ket{A}=0$ (more generally one can use $H_A=H-E_A$) and we neglect the recoil energy of the nucleus. The hadronic tensor is given by 
\bigskip
\begin{equation}
    \label{Hmunu-naive}
    H_{\mu\nu}(q_1,q_2) = \left\langle J_\mu(\vb{q}_1) \frac{1}{\omega_2-H+\iu 0} J_\nu(\vb{q}_2) \right\rangle  + \left\langle J_\nu(\vb{q}_2) \frac{1}{\omega_1-H+\iu 0} J_\mu(\vb{q}_1) \right\rangle + \left\langle S_{\mu\nu}(\vb{q}_1,\vb{q}_2) \right\rangle ~,
\end{equation}
where $S_{\mu\nu}$ is a ``seagull'' vertex. For example, one obtains from minimal coupling of non-relativistic particles $S_{\mu\nu}(\vb{q}_1,\vb{q}_2)= -\frac{1}{M} g_{\perp\mu\nu}\rho(\vb{q}_1+\vb{q}_2)$ where $g_{\perp\mu\nu} = g_{\mu\nu}-v_\mu v_\nu$, $v_\mu = (1,\vb*{0})$ is the four-velocity of the initial nucleus, and $M$ is the constituent particle (e.g., proton) mass. 

As written in \cref{Hmunu-naive}, the energy transfers $\omega_1$ and $\omega_2$ are naively independent variables, however they satisfy $\omega_1+\omega_2=0$ due to the delta function in \cref{M_NLO}. In what follows it is always understood that $\omega=\omega_1=-\omega_2$ and that $H_{\mu\nu}$ is a function of a single independent variable $\omega$, which is important for establishing gauge invariance.

Before proceeding let us review some identities that will be useful in what follows. First, consider the continuity equation $\partial^\mu J_\mu(x) = \partial_t \rho(\vb{x}) + \vb*{\nabla}\cdot \vb{J}(\vb{x}) = 0$ where we have evaluated at $t=0$. Writing the time-derivative using a commutator with the Hamiltonian, and working in momentum space, this becomes 
\begin{equation}
    \label{cont-eq}
    [H,\rho(\vb{q})] = \vb{q} \cdot \vb{J}(\vb{q}) ~.
\end{equation}
The continuity equation implies Siegert's theorem \cite{PhysRev.52.787} (recall we are neglecting the target's recoil energy)
\begin{equation}
    \label{Siegert}
    \mel{A(\vb{p}')}{\vb{q}\cdot \vb{J}(\vb{q})}{A(\vb{p})} =  \mel{A(\vb{p}')}{[H,\rho(\vb{q})]}{A(\vb{p})} = 0~. 
\end{equation}
\Cref{cont-eq} can also be used to show that 
\begin{equation}
    \label{anti-symm-trick}
    \langle \rho(\vb{q}_1)(\vb{q}_2\cdot \vb{J})\rangle = - \langle (\vb{q}_1\cdot \vb{J})\rho(\vb{q}_2)\rangle~,
\end{equation}
where we have used $H\ket{A}=0$. 

\vfill 
\pagebreak 

\subsection{Approximations}
The tensor $H_{\mu\nu}(q_1,q_2)$ must satisfy the Ward identities $q_1^\mu H_{\mu\nu}(q_1, q_2) = q_2^\nu H_{\mu\nu}(q_1,q_2) = 0$. Simple and  physically motivated approximations for evaluating $H_{\mu\nu}(q_1,q_2)$, however, often spoil gauge invariance. For example consider just the ``elastic contribution'' from the same nuclear state $\ketbra{A}$ between the currents,\!\footnote{In the large target-mass limit, the averages $\langle \ldots \rangle$ depend only on the three momenta inserted. }
\begin{equation}
    H_{\mu\nu}^{\rm elastic}(q_1,q_2) = (-2\pi \iu ) \delta(\omega)  \langle J_\mu(\vb{q}_1) \rangle  \langle J_\nu(\vb{q}_2) \rangle  
    + \left\langle S_{\mu\nu}(\vb{q}_1,\vb{q}_2) \right\rangle ~, 
\end{equation}
where we have used a non-relativistic one-body seagull term and imposed $\omega_1=\omega=-\omega_2$. This approximation is not gauge invariant,
\begin{equation}
    q_1^\mu H_{\mu\nu}^{\rm elastic}(q_1,q_2)  =   q_{1}^{\nu} \left\langle S_{\mu\nu}(\vb{q}_1,\vb{q}_2) \right\rangle  ~,
\end{equation}
where we have used \cref{Siegert}. 

Another common approximation scheme is the closure approximation \cite{Friar:1974bn,Jakubassa-Amundsen:2022ckf,Jakubassa-Amundsen:2023wuz,Jakubassa-Amundsen:2024dai} where one replaces $H\rightarrow \bar{E}$ in the definition of \cref{Hmunu-naive}. In this case one obtains (leaving $+\iu 0$ implicit), 
\begin{equation}
    H_{\mu\nu}^{\rm closure}(q_1,q_2) = \frac{1}{-\omega-\bar{E}}\left\langle J_\mu(\vb{q}_1) J_\nu(\vb{q}_2) \right\rangle  + \frac{1}{\omega-\bar{E}} \left\langle J_\nu(\vb{q}_2)  J_\mu(\vb{q}_1) \right\rangle  + \left\langle S_{\mu\nu}(\vb{q}_1,\vb{q}_2) \right\rangle ~.
\end{equation}
This expression does not satisfy the relevant Ward identities. For example, using \cref{cont-eq},
\begin{equation}
q_1^{\mu}H_{\mu \nu}^{\rm closure}=\left\langle \rho ({\bf q}_1) \qty[\frac{\omega+H}{-\omega-\bar{E}}] J_\nu({\bf q}_2)\right\rangle + \left\langle J_\nu({\bf q}_2) \qty[\frac{\omega-H}{\omega-\bar{E}}] \rho ({\bf q}_1)\right\rangle +q_1^{\mu} \left\langle S_{\mu \nu}  \right\rangle\neq 0~ .
\end{equation}

Another common approximation stems from the observation that matrix elements of $\rho$ are large compared to matrix elements of $\vb{J}$. Therefore, in the literature sometimes only the charge density pieces are kept \cite{Friar:1974bn,Jakubassa-Amundsen:2023wuz,Jakubassa-Amundsen:2022ckf}, 
\begin{equation}    
    \label{Hmunu_rhorho}
    H_{\mu\nu}^{\rho\rho} = v_\mu v_\nu \left\langle    \rho(\vb{q}_1) \frac{1}{-\omega -H +\iu 0} \rho(\vb{q}_2)  + \rho(\vb{q}_2) \frac{1}{\omega -H +\iu 0} \rho(\vb{q}_1)\right\rangle~,
\end{equation}
which gives $q_1^\mu H_{\mu\nu}^{\rho\rho}\neq 0$ due to the fact that $[H,\rho(\vb{q})]\neq 0$. Clearly a representation of $H_{\mu\nu}$ from which gauge invariant approximation schemes can be easily derived is highly desirable.

\subsection{Gauge invariance}
The above discussion naturally raises the question: how does one systematically approximate $H_{\mu\nu}$ without spoiling gauge invariance? This is an important question in systems where matrix elements must be evaluated numerically.

One can make substantial progress by making use of the continuity equation \cref{cont-eq}. Let us decompose the current operator in covariant notation as 
\begin{equation}
    J_\mu = \rho v_\mu + J_{\perp \mu}~.
\end{equation}
Then, using \cref{cont-eq} and $H\ket{A}=0$, one can show that 
\begin{equation}
    \label{trick-1}
    \begin{split}
     &\frac{1}{\omega-H+\iu 0}J_\mu(\vb{q}) \ket{A}=\qty[\frac{1}{\omega +\iu 0}\rho(\vb{q})v_\mu +  \frac{1}{\omega-H +\iu 0}\mathscr{J}_{\perp\mu}^{(+)}(q)]\ket{A}  ~,
    \end{split}
\end{equation}
and 
\begin{equation}
    \label{trick-2}
    \begin{split}
     &\bra{A}J_\mu(\vb{q})\frac{1}{-\omega-H +\iu 0}=\bra{A}\qty[\rho(\vb{q})v_\mu\frac{1}{-\omega +\iu 0} +  \mathscr{J}_\mu^{(-)}(q)\frac{1}{-\omega-H +\iu 0}]  ~,
    \end{split}
\end{equation}
where we introduce the operators 
\begin{equation}
    \label{curly-J-def}
    \mathscr{J}_\mu^{(\pm)}(q) = J_{\perp\mu}(\vb{q}) \pm \frac{1}{\pm\omega+\iu 0}(\vb{q}\cdot \vb{J})  v_\mu~.
\end{equation}
These identities are useful because $q^\mu \mathscr{J}_\mu^{(\pm)}(q) =0$ where $\omega=q_0$.

Repeated application leads to one of our main results for elastic electron scattering
    \begin{equation}
        \label{Hmunu-improved}
        \begin{split}
        H_{\mu\nu}(q_1,q_2)& = (-2\pi \iu )\delta(\omega_2) v_\mu v_\nu \langle \rho(\vb{q}_2) \rho(\vb{q}_1) \rangle  \\
        &~~~+ \bigg(~~~\frac{1}{\omega_2} v_\nu \langle J_{\perp\mu}(\vb{q}_1) \rho (\vb{q}_2)  \rangle  + \frac{1}{\omega_1} v_\mu  \langle J_{\perp\nu}(\vb{q}_2) \rho (\vb{q}_1)  \rangle \\
        &\hspace{0.05\linewidth}~+ \frac{1}{-\omega_2} v_\nu \langle \rho(\vb{q}_2) J_{\perp\mu}(\vb{q}_1)  \rangle  + \frac{1}{-\omega_1} v_\mu  \langle  \rho (\vb{q}_1) J_{\perp\nu}(\vb{q}_2)  \rangle~ \\
        &\hspace{0.05\linewidth}~+ \frac{1}{-\omega_2} \frac{1}{\omega_1} v_\mu v_\nu \langle \rho(\vb{q}_2) (\vb{q}_1\cdot \vb{J}) \rangle  + \frac{1}{-\omega_1} \frac{1}{\omega_2} v_\mu v_\nu  \langle  \rho (\vb{q}_1) (\vb{q}_2\cdot \vb{J}) \rangle \\
        &\hspace{0.05\linewidth}~+  \langle S_{\mu\nu} (\vb{q}_1,\vb{q}_2) \rangle  \bigg) \\
        &+ \bigg[ \left\langle \mathscr{J}^{(-)}_\nu (q_2) \frac{1}{\omega_1-H} \mathscr{J}_\mu^{(+)}(q_1) \right\rangle  
            +  \left\langle  \mathscr{J}_\mu^{(-)}(q_1)  \frac{1}{\omega_2-H} \mathscr{J}_\nu^{(+)}(q_2)\right\rangle \bigg] ~,
    \end{split}
    \end{equation}
where we have left all factors of $+\iu 0$ implicit. Notice that, due to the absence of target recoil, this representation contains two terms which are manifestly gauge invariant (upon including $(2\pi)\delta(\omega_1+\omega_2)$ from the energy-conserving delta function), and the term in large parentheses in \cref{Hmunu-improved} which we denote by $(\ldots)_{\mu\nu}$. 

Despite appearances $(\ldots)_{\mu\nu}$ must also be gauge invariant. Contracting with $q_1^\mu$ we obtain, 
\begin{equation}
    \label{non-trivial-GI}
    q_1^\mu (\ldots)_{\mu\nu} =  \left\langle \qty[J_{\perp\nu}(\vb{q}_2) ,\rho (\vb{q}_1) ] \right\rangle +  q_1^{\nu}  \langle S_{\mu\nu}(\vb{q}_1,\vb{q}_2)\rangle ~, 
\end{equation}
where we have used \cref{anti-symm-trick}. For the seagull term    $S_{\mu\nu} = -(g_{\perp\mu\nu}/M)\rho(\vb{q}_1+\vb{q}_2)$, and leading-order one body currents,\!\footnote{Recall that we use a convention where $\rho(\vb{q})$ and $\vb{J}(\vb{q})$ inject momentum $\vb{q}$.}
\begin{equation}
    \label{one-body-currents}
    \begin{split}
        \rho_{\rm 1B}(\vb{q}) &= \int\!\frac{\dd^3 p}{(2\pi)^3} a_{\vb{p}+\vb{q}}^\dagger a_{\vb{p}}~,\\
        \vb{J}_{\rm 1B}(\vb{q}) &= \int\!\frac{\dd^3 p}{(2\pi)^3}  \qty(\frac{2\vb{p}+\vb{q}}{2M}) a_{\vb{p}+\vb{q}}^\dagger a_{\vb{p}}~,
    \end{split}
\end{equation}
with $a_{\vb{p}}$ the proton annihilation operator,  the right-hand side of \cref{non-trivial-GI} can be shown to vanish by explicitly evaluating the commutator,
\begin{equation}
    [\vb{J}_{\rm 1B}(\vb{q}_2),\rho_{\rm 1B}(\vb{q}_1)] = \frac{\vb{q}_1}{M} \rho_{\rm 1B}(\vb{q}_1+\vb{q}_2)~.
\end{equation}
More generally for generic current operators satisfying the continuity equation (e.g, those derived using chiral perturbation theory \cite{Pastore:2008ui,Pastore:2009is,Pastore:2011ip,Kolling:2011mt,Baroni:2021vrc}), gauge invariance demands that the Schwinger term  \cite{Schwinger:1959xd,Boulware:1966qsd,Brown:1966zza} (the left-hand side) is related to the seagull vertex (the right-hand side),
\begin{equation}
    \label{schwinger-term}
    \left\langle \qty[J_{\perp\nu}(\vb{q}_2) ,\rho (\vb{q}_1) ] \right\rangle  =
    -  q_1^\mu \langle S_{\mu\nu}(\vb{q}_1,\vb{q}_2) \rangle   ~. 
\end{equation}

The commutator in \cref{schwinger-term} allows us to further decompose $(\ldots)_{\mu\nu}=\{\ldots\}_{\mu\nu}+ [\ldots]_{\mu\nu}$ into a piece containing only anti-commutators $\{\ldots\}_{\mu\nu}$ and a piece containing only commutators $[\ldots]_{\mu\nu}$, the latter of which can the be written in terms of $\langle S_{\mu\nu} \rangle$, 
\bigskip
\begin{align}
   \begin{split}
    \label{anti-commutator-piece}
       \{\ldots\}_{\mu\nu}&= 
     (-\iu\pi) \delta(\omega) \langle \{J_{\perp,\nu}(\vb{q}_2),\rho(\vb{q}_1)\}\rangle  v_\mu
   + (-\iu\pi) \delta(\omega) \langle \{J_{\perp,\mu}(\vb{q}_1),\rho(\vb{q}_2)\}\rangle v_\nu
   \\[1pt]
   &\hspace{0.15\linewidth} + (-\iu\pi)\delta'(\omega) v_{\mu}v_\nu  
    \big( \left\langle \rho(\vb{q}_2)H \rho(\vb{q}_1)\right\rangle 
    - \left\langle \rho(\vb{q}_1)H \rho(\vb{q}_2) \right\rangle \big) ~.
   \end{split}~\\[12pt]
    \begin{split}
        \label{commutator-piece}
   [\ldots]_{\mu\nu} &= -{\rm PV}\qty(\frac{1}{\omega_1}) v_\mu q_{1}^\alpha\langle S_{\alpha\nu} \rangle - {\rm PV}\qty(\frac{1}{\omega_2})v_\nu q_{2}^\alpha\langle S_{\alpha\mu} \rangle  \\
   &\hspace{0.2\linewidth}- \frac{1}{2}\qty(\frac{1}{\omega_1} \frac{1}{-\omega_2}  + \frac{1}{\omega_2} \frac{1}{-\omega_1} ) v_\mu v_\nu q_1^{\alpha}q_2^{\beta}\langle S_{\alpha\beta} \rangle ~,
   \end{split}
\end{align}
Notice that $q^\mu_1\{\ldots\}_{\mu\nu}=q^\nu_2 \{\ldots\}_{\mu\nu}=0$. When using one-body currents as defined in \cref{one-body-currents} and the associated seagull operator, $S_{\mu\nu} = - g_{\perp\mu\nu} \rho(\vb{q}_1+\vb{q}_2)/M$, one finds 
\begin{equation}
    \begin{split}
    [\ldots]_{\mu\nu} &= \frac{1}{M}\bigg[{\rm PV}\qty(\frac{1}{\omega_1}) v_\mu q_{1\perp,\nu}+ {\rm PV}\qty(\frac{1}{\omega_2})v_\nu q_{2\perp,\mu} \\
    &\hspace{0.1\linewidth}+ \frac{1}{2}\qty(\frac{1}{\omega_1} \frac{1}{-\omega_2}  + \frac{1}{\omega_2} \frac{1}{-\omega_1} ) v_\mu v_\nu (q_{1\perp} \!\cdot q_{2\perp}) \bigg] \langle \rho(\vb{q}_1+\vb{q}_2)\rangle~.
    \end{split}
\end{equation}
%
Using \cref{commutator-piece,anti-commutator-piece} we may now arrive at our main result in fully simplified form, by rewriting \cref{Hmunu-improved} as 
    \begin{equation}
        \label{Hmunu-improved++}
        \begin{split}
        H_{\mu\nu}(q_1,q_2)& = (-2\pi \iu )\delta(\omega_2) v_\mu v_\nu \langle \rho(\vb{q}_2) \rho(\vb{q}_1) \rangle + \{\ldots \}_{\mu\nu} + \big([\ldots ]_{\mu\nu}+  \langle S_{\mu\nu} (\vb{q}_1,\vb{q}_2) \rangle \big) \\
        &+ \bigg[ \left\langle \mathscr{J}^{(-)}_\nu (q_2) \frac{1}{\omega_1-H} \mathscr{J}_\mu^{(+)}(q_1) \right\rangle  
            +  \left\langle  \mathscr{J}_\mu^{(-)}(q_1)  \frac{1}{\omega_2-H} \mathscr{J}_\nu^{(+)}(q_2)\right\rangle \bigg] ~.
    \end{split}
    \end{equation}
Notice that \cref{Hmunu-improved++} is composed of four separately gauge invariant terms that are organized in an expansion in $1/M$ for the one-body currents as given in \cref{one-body-currents}, 
\begin{enumerate}
    \item The $\langle \rho \rho \rangle$ term begins at $O(1)$, and has no Greens function. 
    \item The $\{\ldots \}_{\mu\nu}$ begins at $O(1/M)$ and also has no Greens function. 
    \item The $[\ldots ]_{\mu\nu} + S_{\mu\nu}$ term begins at $O(1/M)$ and depends only on $\langle S_{\mu\nu}\rangle$. For the one body seagull term this is just the elastic form $\langle \rho\rangle$. 
    \item The term involving $\mathscr{J}\!\!\!\mathscr{J}$ begins at $O(1/M^2)$ and includes a nuclear Greens function. Since $q^\mu \mathscr{J}_\mu(q)=0$ this term remains gauge invariant even if the Greens function is approximated. 
\end{enumerate}
\subsection{Connection to Compton scattering}
The tensor $H_{\mu\nu}(q_1,q_2)$ contains Lorentz structures that are typically omitted in the discussion of the (closely related) Compton tensor, see e.g., Refs \cite{Tarrach:1975tu,Hutt:1999pz,Gorchtein:2005za,Hill:2016bjv}. For example the term $v_\mu v_\nu$ would typically be argued to be forbidden since this does not satisfy current conservation $q^\mu v_\mu v_\nu = \omega v_\nu$. The key difference between $H_{\mu\nu}(q_1,q_2)$ and a conventional Compton tensor is that $H_{\mu\nu}(q_1,q_2)$ is distribution-valued. With the inclusion of the relevant delta functions, the Lorentz structure $v_\mu v_\nu \delta(\omega_1) \delta(\omega_2)$ is gauge invariant. 

When considering the Compton scattering of real photons, $\omega_1$ and $\omega_2$ are non-vanishing and all terms proportional to $\delta(\omega)$ disappear from the amplitude. We then find for $\omega\neq0$,
\begin{equation}
    H_{\mu\nu} = [\ldots]_{\mu\nu} + \langle S_{\mu\nu} \rangle +    \left\langle \mathscr{J}^{(-)}_\nu (q_2) \frac{1}{\omega_1-H} \mathscr{J}_\mu^{(+)}(q_1) +  \mathscr{J}_\mu^{(-)}(q_1)  \frac{1}{\omega_2-H} \mathscr{J}_\nu^{(+)}(q_2)\right\rangle ~.
\end{equation}
%

Returning to the case of electron scattering, although $H_{\mu\nu}(q_1,q_2)$ is distribution valued, the matrix element $\mathcal{M}^{(1)}$ as given in \cref{M_NLO} is not, because an integral over $\omega$ is performed. The integral over $\omega$ can be performed by standard contour methods. Care must be taken to properly account for double poles and/or triple poles which appear in the $\mathscr{J}\!\!\!\mathscr{J}$ and $(\ldots)_{\mu\nu}$ terms. The remaining expressions then contain only nuclear matrix elements of currents at definite three-momentum $\vb{q}_1$ and $\vb{q}_2=\vb{Q}-\vb{q}_1$, which must be integrated over using $\int \dd^3 q_1(\ldots )$. 

\subsection{Gauge invariant approximations}
In the above discussion we emphasized that sensible approximation schemes can spoil gauge invariance when applied directly to \cref{Hmunu-naive}. The merit of \cref{Hmunu-improved,Hmunu-improved++} is that the only term with a nuclear Greens function (the $\mathscr{J}\!\!\!\mathscr{J}$ term) is manifestly gauge invariant in {\it any} approximation scheme. The $\rho\rho$ and the $\{ ...  \}_{\mu\nu}$ term contains no Greens function, and can therefore be evaluated given a nuclear or atomic wavefunction for $\ket{A}$. As we discuss below, \cref{Hmunu-improved++} is particularly well suited to approximations. 

As an example consider the closure approximation. The $\mathscr{J}\!\!\!\mathscr{J}$terms (now keeping $\bar{E}$) can be approximated by 
\begin{equation}
    \bigg[ \frac{1}{\omega-\bar{E}+\iu 0} \left\langle \mathscr{J}^{(-)}_\nu (q_2) \mathscr{J}_\mu^{(+)}(q_1) \right\rangle  
            +  \frac{1}{-\omega-\bar{E}+\iu 0} \left\langle  \mathscr{J}_\mu^{(-)}(q_1) \mathscr{J}_\nu^{(+)}(q_2)\right\rangle \bigg]~,
\end{equation}
which manifestly satisfies the necessary Ward identities since $q^\mu \mathscr{J}_\mu (q)=0$. 
The value of $\bar{E}$ may be chosen based on whatever matrix elements are expected to dominate the sum. When combined with the other terms defined in \cref{Hmunu-improved}, this approximation provides a gauge invariant improved closure approximation. 

Next consider the elastic approximation where $\ketbra{A}$ is inserted between operators. Gauge invariance is immediately clear for all the terms except $\{~~~\}_{\mu \nu}$
where $q_1^{\mu}\{~~~\}_{\mu \nu}=0$  follows from \cref{cont-eq} and $H \ket{A} =0$. In fact the Ward identities hold even for the insertion of an arbitrary state $\ketbra{B}$ between the density and current operators not just the ground state (even though $H\ket{B} \neq 0$). Doing this insertion,  using again  
\cref{cont-eq} and $H\ket{A}=0$ as well as $\delta'(\omega) \omega =-\delta(\omega)$ it is easy to verify gauge invariance. This means that improving the approximation is straightforward and systematic; any set of intermediate states can be included until results stabilize numerically.


For non-relativistic bound states, with momentum injected by the currents satisfying $|\vb{q}|\ll M$,  \cref{Hmunu-improved} is organized such that each of the three gauge invariant contributions begins at one higher order as an expansion in $1/M$ (as already emphasized below \cref{Hmunu-improved++}). The $\rho\rho$ term begins at $O(1/M^0)$, the 
$[ \ldots ]_{\mu \nu}$ and $\{ \ldots \}_{\mu \nu}$ terms begin at $O(1/M^1)$, and the $\mathscr{J}\!\!\!\mathscr{J}$ term begins at $O(1/M^2)$.  The dominant $\rho\rho$ term is both separately gauge invariant and supplies the dominant contribution to $H_{\mu\nu}$. 

In addition to the organization of the amplitude in powers of $1/M$, coherent enhancements proportional to the number of protons, $Z$, in the nucleus lead to the hierarchy $\langle \rho(\vb{q}) \rangle\gg \langle \vb{J}(\vb{q})\rangle$.  For this reason the $\rho\rho$ term dominates the cross section since it is both coherent, and unsuppressed in the heavy constituent mass limit where $M\rightarrow \infty$. 

\section{Weak interaction processes}
Having discussed electron scattering, let us now apply the same techniques to processes mediated by the weak interaction. Unlike electron scattering, where we consider two insertions of the electromagnetic current, for electromagnetic corrections to weak interaction processes we only have one electromagnetic current insertion. We will denote any weak operator by $\mathcal{O}_W$. 

Let us consider a hadronic correlator with one electromagnetic current insertion. Since we consider a different initial, $\ket{A}$, and final state, $\ket{B}$, we introduce $H_A=H-E_A$ and $H_B=H-E_B$, where in the above discussion we have set $E_A=0$. In terms of $H_A$ and $H_B$ we have, 
\begin{equation}
    \label{weak-correlator-def}
    H_\mu = 
    \left\langle J_\mu(\vb{q}) \frac{1}{-\omega-H_B+\iu 0} \mathcal{O}_W(\vb{k})  \right\rangle + 
    \left\langle  \mathcal{O}_W(\vb{k}) \frac{1}{\omega-H_A+\iu 0} J_\mu(\vb{q}) \right\rangle + \left\langle \mathcal{S}_{\mu}(\vb{q},\vb{k}) \right\rangle ~.
\end{equation}
Where the weak analog of the seagull vertex, $\mathcal{S}_{\mu}$, stems from terms in which the photon is attached directly to the weak current. This will occur, for example, whenever covariant derivative couplings are present in the operator $\mathcal{O}_W$.

Applying \cref{trick-1,trick-2}, we obtain 
\begin{equation}
    \begin{split}
    H_\mu = 
    &v_\mu \qty(\left\langle \rho(\vb{q}) \frac{1}{-\omega+\iu 0} \mathcal{O}_W(\vb{k})  \right\rangle + 
    \left\langle  \mathcal{O}_W(\vb{k}) \frac{1}{\omega+\iu 0} \rho(\vb{q}) \right\rangle ) \\
    &+   \left\langle \mathscr{J}_\mu^{(-)}(\vb{q}) \frac{1}{-\omega-H_B+\iu 0} \mathcal{O}_W(\vb{k})  \right\rangle + 
    \left\langle  \mathcal{O}_W(\vb{k}) \frac{1}{\omega-H_A+\iu 0} \mathscr{J}_\mu^{(+)}(\vb{q}) \right\rangle + \left\langle \mathcal{S}_{\mu}(\vb{q},\vb{k}) \right\rangle ~.
    \end{split}
\end{equation}
This can be re-written as 
\begin{equation}
    \label{weak-correlator}
    \begin{split}
    H_\mu = 
    &v_\mu (-2\pi \iu )\delta(\omega)\left\langle \rho(\vb{q})   \mathcal{O}_W(\vb{k})  \right\rangle + 
    \frac{v_\mu}{\omega+\iu 0}\left\langle  \qty[\mathcal{O}_W(\vb{k}),  \rho(\vb{q})] \right\rangle  \\
    &+   \left\langle \mathscr{J}_\mu^{(-)}(\vb{q}) \frac{1}{-\omega-H_B+\iu 0} \mathcal{O}_W(\vb{k})  \right\rangle + 
    \left\langle  \mathcal{O}_W(\vb{k}) \frac{1}{\omega-H_A+\iu 0} \mathscr{J}_\mu^{(+)}(\vb{q}) \right\rangle + \left\langle \mathcal{S}_{\mu}(\vb{q},\vb{k}) \right\rangle ~.
    \end{split}
\end{equation}
Notice that when $\mathcal{O}_W$ commutes with $\rho$ the hadronic correlator is gauge invariant. When considering the charged weak current (e.g., in beta decays, neutrino scattering, or muon capture) $\mathcal{O}_W$ will {\it not} commute with $\rho$.

\subsection{Radiative muon capture}
As an example, let us consider the leading one-body contribution to time-like component of the weak charged current, 
\begin{equation}
    \mathcal{O}_W^{(0)}(\vb{k})=\int \frac{\dd^3p}{(2\pi)^3} b^\dagger_{\vb{p}+\vb{k}} a_{\vb{p}} ~, 
\end{equation}
where $b^\dagger_{\vb{p}+\vb{k}}$ is a neutron creation operator and $a_{\vb{p}}$ the proton annihilation operator. Since $\mathcal{O}_W$ does not contain covariant derivatives we have $\mathcal{S}_{\mu}=0$. As a concrete physical application, consider radiative muon capture at threshold,\!\footnote{We define the photon momenta such that $\vb{q}$ is routed into (rather than out of) the nuclear system. } 
\begin{equation}
    \mu(\vb{0}) A(\vb{0}) \rightarrow \nu(-\vb{k}) \gamma(-\vb{q}) B(\vb{Q})~,
\end{equation}
where $\vb{k}+ \vb{p}=\vb{Q}$. In a general gauge there are two classes of diagrams that contribute to this process at tree level: $\mathcal{M}_{a}$ where the photon is radiated off the muon line, and $\mathcal{M}_{b}$ where the photon is radiated off the nucleus 
\bigskip
\begin{equation}
    \mathcal{M}_a= 
     \raisebox{-20pt}{
    \begin{fmffile}{matrix-el-a}
    \begin{fmfgraph*}(15,12)
    \fmfcurved
    \fmftop{t1,t2,t3}
    \fmfbottom{b1,b2,b3,b4}
    \fmfleft{i1,i2}
    \fmfright{o1,o2}
    \fmf{double}{i1,v2,o1}
    \fmf{plain}{i2,v2,o2}
    \fmffreeze
    \fmf{phantom}{i2,v1,v2,o2}
    \fmf{photon,tension=0}{v1,t2}
    \fmfv{label=$A$}{i1}
    \fmfv{label=$\gamma$}{t2}
    \fmfv{label=$B$}{o1}
    \fmfv{label=$\mu$}{i2}
    \fmfv{label=$\nu$}{o2}
\end{fmfgraph*}
\end{fmffile} } 
    \quad,\quad
    \mathcal{M}_b= 
     \raisebox{-20pt}{
    \begin{fmffile}{matrix-el-b}
    \begin{fmfgraph*}(15,12) 
    \fmfcurved
    \fmftop{t1,t2,t3}
    \fmfbottom{b1,b2,b3}
    \fmfleft{i1,i2}
    \fmfright{o1,o2}
    \fmf{double}{i1,v2,o1}
    \fmf{plain}{i2,v2,o2}
    \fmffreeze
    \fmf{phantom}{i1,v1,v2,o1}
    \fmf{photon,tension=0}{v1,b2}
    \fmfv{label=$A$}{i1}
    \fmfv{label=$\gamma$}{b2}
    \fmfv{label=$B$}{o1}
    \fmfv{label=$\mu$}{i2}
    \fmfv{label=$\nu$}{o2}
\end{fmfgraph*}
\end{fmffile} }
\quad+\quad
     \raisebox{-20pt}{
    \begin{fmffile}{matrix-el-c}
    \begin{fmfgraph*}(15,12) 
    \fmfcurved
    \fmftop{t1,t2,t3}
    \fmfbottom{b1,b2,b3}
    \fmfleft{i1,i0,i2}
    \fmfright{o1,o0,o2}
    \fmf{double}{i1,v2,o1}
    \fmf{plain}{i2,v2,o2}
    \fmffreeze
    \fmf{phantom}{i1,v2,v1,o1}
    \fmf{photon,tension=0}{v1,o0}
    \fmfv{label=$A$}{i1}
    \fmfv{label=$\gamma$}{o0}
    \fmfv{label=$B$}{o1}
    \fmfv{label=$\mu$}{i2}
    \fmfv{label=$\nu$}{o2}
\end{fmfgraph*}
\end{fmffile} }
\quad.\\[15pt]
\end{equation}
The term $\mathcal{M}_{a}$ involves the tree-level hadronic matrix element $\langle\mathcal{O}_W(\vb{Q}) \rangle$; taken alone $\mathcal{M}_a$ is not gauge invariant. The second set of terms $\mathcal{M}_b$ involves a hadronic correlator of the form discussed in \cref{weak-correlator}; these terms are also not gauge invariant in isolation. Since we consider a real photon, we can drop all terms proportional to $\delta(\omega)$ and obtain for hadronic correlator $H_{b,\mu}$ in $\mathcal{M}_b$, 
\begin{equation} \label{RMC-correlator}
    \begin{split}
    H_{b,\mu}= &v_\mu \frac{1}{\omega+\iu 0} \left\langle  [\mathcal{O}_W^{(0)}(\vb{k}),\rho(\vb{q})]  \right\rangle \\
    &\hspace{0.05\linewidth}+   \left\langle \mathscr{J}_\mu^{(-)}(\vb{q}) \frac{1}{-\omega-H_B+\iu 0} \mathcal{O}_W^{(0)}(\vb{k})  \right\rangle + 
    \left\langle  \mathcal{O}_W^{(0)}(\vb{k})\frac{1}{\omega-H_A+\iu 0} \mathscr{J}_\mu^{(+)}(\vb{q}) \right\rangle~,
    \end{split}
\end{equation}
where angle brackets now involve matrix elements of the form $\mel{B(\vb{Q})}{\ldots}{A(\vb{0})}$. Explicitly evaluating the commutator with one-body currents, we find 
\begin{equation}
    \label{RMC-commutator}
    [\mathcal{O}_W^{(0)}(\vb{k}),\rho_{\rm 1B}(\vb{q})] = \mathcal{O}_W^{(0)}(\vb{q}+\vb{k})~. 
\end{equation}
This term is precisely what is required to ensure that $\mathcal{M}_a+\mathcal{M}_b$ is gauge invariant. The terms involving $\mathscr{J}^{\mu}$ are automatically gauge invariant. 

Similar commutators are important to ensure gauge invariance when the photon is virtual rather than real (connecting to the charged lepton line). For example in a beta decay of the form, $A(\vb{0}) \rightarrow B(\vb{Q}) \nu_e(\vb{k}) e^+ (\vb{p})$, the resultant loop amplitude that includes the propagator $1/(\omega+\iu 0)$ becomes proportional to $\left\langle \mathcal{O}_W^{(0)}(\vb{Q}) \right\rangle$. This is required for gauge invariance since the gauge-dependent wavefunction renormalization multiplies the tree-level amplitude which is proportional to  $\left\langle\mathcal{O}_W^{(0)}(\vb{Q})\right\rangle$.

\section{Conclusions}
Electromagnetic radiative corrections are increasingly important for interpreting precision observables involving atoms, nuclei, and other composite particles \cite{Ent:2001hm,Hill:2010yb,Tomalak:2020zfh,Lovato:2020kba,Rocco:2020jlx,Tomalak:2021hec,Tomalak:2022xup,Ruso:2022qes,Seng:2022cnq,Hill:2023acw,Hill:2023bfh,Cirigliano:2023fnz,Cirigliano:2024nfi,Cirigliano:2024rfk,Borah:2024ghn,VanderGriend:2025mdc,Afanasev:2023gev,Richardson:2023vyf,Seng:2024zuc,Combes:2024pvm}. At one-loop order new sensitivity to the inelastic excitations of the bound states emerge, and the proper implementation of gauge invariance requires care. Seemingly innocuous approximation schemes can spoil gauge invariance. 

We have shown how \cref{trick-1,trick-2,curly-J-def}, along with the equal time commutators as given in \cref{schwinger-term}, allows for a manifestly gauge invariant representation of two-photon exchange graphs that appear in the theory of electron scattering off nuclei and atoms. We have applied the same ideas to the weak interaction, and found new representations of the amplitude which make gauge invariance manifestly apparent. Our results are relevant for nuclear beta decays \cite{Seng:2018qru,Hardy:2020qwl,Cirigliano:2024rfk,Hill:2023acw}, parity violating electron scattering \cite{CREX:2022kgg,PREX:2021umo}, neutrino nucleus scattering \cite{Lovato:2020kba,Rocco:2020jlx,Tomalak:2021hec,Tomalak:2022xup,Ruso:2022qes}, and  rare nuclear processes that must be controlled to percent-level precision.

In the case of electron scattering, our results are most relevant for the ``dispersive correction'' \cite{Bethe:1971es,Friar:1974bn,Offermann:1991ft,Jakubassa-Amundsen:2022ckf,Jakubassa-Amundsen:2024dai}. At a numerical level, the impact of our work can be estimated by comparing to Figs.\ 6-10 of Ref.~\cite{Friar:1974bn} for $^{16}{\rm O}$ and $^{12}{\rm C}$. In that work $H_{\mu\nu}$ is approximated by $H_{\mu\nu}^{\rho\rho}$ and evaluated in the closure approximation, as given in \cref{Hmunu_rhorho},  with a finite $\bar{E}$. The amplitude $\mathcal{M}^{(1)}$ is then evaluated with $H_{\mu\nu}^{\rho\rho}$ using Coulomb gauge photon propagators. Therefore, according to our \cref{Hmunu-improved,Hmunu-improved++}, their $\bar{E}=0$ calculation represents a gauge-invariant approximation. By way of contrast, for finite values of $\bar{E}$ the resultant correction is gauge dependent. The differences between the gauge dependent part gives an overall $\sim1\%$ correction to the cross section \cite{Friar:1974bn}; this is a sizeable error relative to the few-percent dispersive correction itself. In the case of the weak interaction \cref{RMC-correlator,RMC-commutator} show how, at one-loop level, gauge invariance is enforced for charged currents where one must deal simultaneously with the leptonic and hadronic parts of the diagram. 

While we have emphasized the approximation of nuclear/atomic matrix elements, our approach is also useful when the Hamiltonian and current operators are approximate rather than exact \cite{Pastore:2008ui,Pastore:2009is,Pastore:2011ip,Epelbaum:2019kcf,Friman-Gayer:2020vqn,Baroni:2021vrc,Cirigliano:2018yza,Cirigliano:2018yza,Cirigliano:2024rfk}. In this case the continuity equation \cref{cont-eq} will only be approximately satisfied, and for purposes of gauge invariance higher order terms should be consistently neglected. After manipulating the amplitude into the representations \cref{Hmunu-improved,Hmunu-improved++}, we are guaranteed that when the resulting expression is evaluated with approximate currents it will satisfy the Ward identity {\it exactly}. 

Dispersive corrections are notoriously difficult to handle \cite{Friar:1974bn,Afanasev:2005mp,Gorchtein:2011mz,Arrington:2011dn,Seng:2018qru,Pasquini:2018wbl,JeffersonLabHallA:2018cmf,Kutz:2022hka,Jakubassa-Amundsen:2023wuz,Jakubassa-Amundsen:2024dai,Gennari:2024sbn}, and the representations developed here should allow for new approximation schemes. In addition to the ``box graphs'' discussed in this work, radiative corrections must generally include vertex corrections and wavefunction renormalization on the hadronic lines. The techniques developed above bear on the gauge dependence of these terms, and we plan to pursue this in future work. Including target recoil is another potential improvement.  

The results and techniques developed here should have broad applicability ranging from fundamental physics applications such as neutrino nucleus scattering and measurements of quark mixing with beta decays, to studies of nuclear and/or atomic structure. Future work will apply these ideas in detail to some of the specific examples mentioned above. 

\vfill
\pagebreak

\section*{Acknowledgments} 
\vspace{-12pt}
We thank Bradley Filippone and Gil Paz for helpful feedback. RP and MBW are supported by the U.S. Department of Energy, Office of Science, Office of High Energy Physics under Award Number DE-SC0011632, and by the Walter Burke Institute for Theoretical Physics. RP is supported by the Neutrino Theory Network under Award Number DEAC02-07CH11359.

\bibliographystyle{apsrev4-1}
\bibliography{biblio}

\begin{thebibliography}{55}%
\makeatletter
\providecommand \@ifxundefined [1]{%
 \@ifx{#1\undefined}
}%
\providecommand \@ifnum [1]{%
 \ifnum #1\expandafter \@firstoftwo
 \else \expandafter \@secondoftwo
 \fi
}%
\providecommand \@ifx [1]{%
 \ifx #1\expandafter \@firstoftwo
 \else \expandafter \@secondoftwo
 \fi
}%
\providecommand \natexlab [1]{#1}%
\providecommand \enquote  [1]{``#1''}%
\providecommand \bibnamefont  [1]{#1}%
\providecommand \bibfnamefont [1]{#1}%
\providecommand \citenamefont [1]{#1}%
\providecommand \href@noop [0]{\@secondoftwo}%
\providecommand \href [0]{\begingroup \@sanitize@url \@href}%
\providecommand \@href[1]{\@@startlink{#1}\@@href}%
\providecommand \@@href[1]{\endgroup#1\@@endlink}%
\providecommand \@sanitize@url [0]{\catcode `\\12\catcode `\$12\catcode
  `\&12\catcode `\#12\catcode `\^12\catcode `\_12\catcode `\%12\relax}%
\providecommand \@@startlink[1]{}%
\providecommand \@@endlink[0]{}%
\providecommand \url  [0]{\begingroup\@sanitize@url \@url }%
\providecommand \@url [1]{\endgroup\@href {#1}{\urlprefix }}%
\providecommand \urlprefix  [0]{URL }%
\providecommand \Eprint [0]{\href }%
\providecommand \doibase [0]{http://dx.doi.org/}%
\providecommand \selectlanguage [0]{\@gobble}%
\providecommand \bibinfo  [0]{\@secondoftwo}%
\providecommand \bibfield  [0]{\@secondoftwo}%
\providecommand \translation [1]{[#1]}%
\providecommand \BibitemOpen [0]{}%
\providecommand \bibitemStop [0]{}%
\providecommand \bibitemNoStop [0]{.\EOS\space}%
\providecommand \EOS [0]{\spacefactor3000\relax}%
\providecommand \BibitemShut  [1]{\csname bibitem#1\endcsname}%
\let\auto@bib@innerbib\@empty
\bibitem [{\citenamefont {Ent}\ \emph {et~al.}(2001)\citenamefont {Ent},
  \citenamefont {Filippone}, \citenamefont {Makins}, \citenamefont {Milner},
  \citenamefont {O'Neill},\ and\ \citenamefont {Wasson}}]{Ent:2001hm}%
  \BibitemOpen
  \bibfield  {author} {\bibinfo {author} {\bibfnamefont {R.}~\bibnamefont
  {Ent}}, \bibinfo {author} {\bibfnamefont {B.~W.}\ \bibnamefont {Filippone}},
  \bibinfo {author} {\bibfnamefont {N.~C.~R.}\ \bibnamefont {Makins}}, \bibinfo
  {author} {\bibfnamefont {R.~G.}\ \bibnamefont {Milner}}, \bibinfo {author}
  {\bibfnamefont {T.~G.}\ \bibnamefont {O'Neill}}, \ and\ \bibinfo {author}
  {\bibfnamefont {D.~A.}\ \bibnamefont {Wasson}},\ }\href {\doibase
  10.1103/PhysRevC.64.054610} {\bibfield  {journal} {\bibinfo  {journal} {Phys.
  Rev. C}\ }\textbf {\bibinfo {volume} {64}},\ \bibinfo {pages} {054610}
  (\bibinfo {year} {2001})}\BibitemShut {NoStop}%
\bibitem [{\citenamefont {Hill}\ and\ \citenamefont {Paz}(2010)}]{Hill:2010yb}%
  \BibitemOpen
  \bibfield  {author} {\bibinfo {author} {\bibfnamefont {R.~J.}\ \bibnamefont
  {Hill}}\ and\ \bibinfo {author} {\bibfnamefont {G.}~\bibnamefont {Paz}},\
  }\href {\doibase 10.1103/PhysRevD.82.113005} {\bibfield  {journal} {\bibinfo
  {journal} {Phys. Rev. D}\ }\textbf {\bibinfo {volume} {82}},\ \bibinfo
  {pages} {113005} (\bibinfo {year} {2010})},\ \Eprint
  {http://arxiv.org/abs/1008.4619} {arXiv:1008.4619} \BibitemShut {NoStop}%
\bibitem [{\citenamefont {Tomalak}\ \emph {et~al.}(2021)\citenamefont
  {Tomalak}, \citenamefont {Machado}, \citenamefont {Pandey},\ and\
  \citenamefont {Plestid}}]{Tomalak:2020zfh}%
  \BibitemOpen
  \bibfield  {author} {\bibinfo {author} {\bibfnamefont {O.}~\bibnamefont
  {Tomalak}}, \bibinfo {author} {\bibfnamefont {P.}~\bibnamefont {Machado}},
  \bibinfo {author} {\bibfnamefont {V.}~\bibnamefont {Pandey}}, \ and\ \bibinfo
  {author} {\bibfnamefont {R.}~\bibnamefont {Plestid}},\ }\href {\doibase
  10.1007/JHEP02(2021)097} {\bibfield  {journal} {\bibinfo  {journal} {JHEP}\
  }\textbf {\bibinfo {volume} {02}},\ \bibinfo {pages} {097} (\bibinfo {year}
  {2021})},\ \Eprint {http://arxiv.org/abs/2011.05960} {arXiv:2011.05960}
  \BibitemShut {NoStop}%
\bibitem [{\citenamefont {Rocco}(2020)}]{Rocco:2020jlx}%
  \BibitemOpen
  \bibfield  {author} {\bibinfo {author} {\bibfnamefont {N.}~\bibnamefont
  {Rocco}},\ }\href {\doibase 10.3389/fphy.2020.00116} {\bibfield  {journal}
  {\bibinfo  {journal} {Front. in Phys.}\ }\textbf {\bibinfo {volume} {8}},\
  \bibinfo {pages} {116} (\bibinfo {year} {2020})}\BibitemShut {NoStop}%
\bibitem [{\citenamefont {Lovato}\ \emph {et~al.}(2020)\citenamefont {Lovato},
  \citenamefont {Carlson}, \citenamefont {Gandolfi}, \citenamefont {Rocco},\
  and\ \citenamefont {Schiavilla}}]{Lovato:2020kba}%
  \BibitemOpen
  \bibfield  {author} {\bibinfo {author} {\bibfnamefont {A.}~\bibnamefont
  {Lovato}}, \bibinfo {author} {\bibfnamefont {J.}~\bibnamefont {Carlson}},
  \bibinfo {author} {\bibfnamefont {S.}~\bibnamefont {Gandolfi}}, \bibinfo
  {author} {\bibfnamefont {N.}~\bibnamefont {Rocco}}, \ and\ \bibinfo {author}
  {\bibfnamefont {R.}~\bibnamefont {Schiavilla}},\ }\href {\doibase
  10.1103/PhysRevX.10.031068} {\bibfield  {journal} {\bibinfo  {journal} {Phys.
  Rev. X}\ }\textbf {\bibinfo {volume} {10}},\ \bibinfo {pages} {031068}
  (\bibinfo {year} {2020})},\ \Eprint {http://arxiv.org/abs/2003.07710}
  {arXiv:2003.07710} \BibitemShut {NoStop}%
\bibitem [{\citenamefont {Tomalak}\ \emph
  {et~al.}(2022{\natexlab{a}})\citenamefont {Tomalak}, \citenamefont {Chen},
  \citenamefont {Hill},\ and\ \citenamefont {McFarland}}]{Tomalak:2021hec}%
  \BibitemOpen
  \bibfield  {author} {\bibinfo {author} {\bibfnamefont {O.}~\bibnamefont
  {Tomalak}}, \bibinfo {author} {\bibfnamefont {Q.}~\bibnamefont {Chen}},
  \bibinfo {author} {\bibfnamefont {R.~J.}\ \bibnamefont {Hill}}, \ and\
  \bibinfo {author} {\bibfnamefont {K.~S.}\ \bibnamefont {McFarland}},\ }\href
  {\doibase 10.1038/s41467-022-32974-x} {\bibfield  {journal} {\bibinfo
  {journal} {Nature Commun.}\ }\textbf {\bibinfo {volume} {13}},\ \bibinfo
  {pages} {5286} (\bibinfo {year} {2022}{\natexlab{a}})},\ \Eprint
  {http://arxiv.org/abs/2105.07939} {arXiv:2105.07939} \BibitemShut {NoStop}%
\bibitem [{\citenamefont {Tomalak}\ \emph
  {et~al.}(2022{\natexlab{b}})\citenamefont {Tomalak}, \citenamefont {Chen},
  \citenamefont {Hill}, \citenamefont {McFarland},\ and\ \citenamefont
  {Wret}}]{Tomalak:2022xup}%
  \BibitemOpen
  \bibfield  {author} {\bibinfo {author} {\bibfnamefont {O.}~\bibnamefont
  {Tomalak}}, \bibinfo {author} {\bibfnamefont {Q.}~\bibnamefont {Chen}},
  \bibinfo {author} {\bibfnamefont {R.~J.}\ \bibnamefont {Hill}}, \bibinfo
  {author} {\bibfnamefont {K.~S.}\ \bibnamefont {McFarland}}, \ and\ \bibinfo
  {author} {\bibfnamefont {C.}~\bibnamefont {Wret}},\ }\href {\doibase
  10.1103/PhysRevD.106.093006} {\bibfield  {journal} {\bibinfo  {journal}
  {Phys. Rev. D}\ }\textbf {\bibinfo {volume} {106}},\ \bibinfo {pages}
  {093006} (\bibinfo {year} {2022}{\natexlab{b}})},\ \Eprint
  {http://arxiv.org/abs/2204.11379} {arXiv:2204.11379} \BibitemShut {NoStop}%
\bibitem [{\citenamefont {Seng}\ and\ \citenamefont
  {Gorchtein}(2023)}]{Seng:2022cnq}%
  \BibitemOpen
  \bibfield  {author} {\bibinfo {author} {\bibfnamefont {C.-Y.}\ \bibnamefont
  {Seng}}\ and\ \bibinfo {author} {\bibfnamefont {M.}~\bibnamefont
  {Gorchtein}},\ }\href {\doibase 10.1103/PhysRevC.107.035503} {\bibfield
  {journal} {\bibinfo  {journal} {Phys. Rev. C}\ }\textbf {\bibinfo {volume}
  {107}},\ \bibinfo {pages} {035503} (\bibinfo {year} {2023})},\ \Eprint
  {http://arxiv.org/abs/2211.10214} {arXiv:2211.10214} \BibitemShut {NoStop}%
\bibitem [{\citenamefont {Ruso}\ \emph {et~al.}(2025)\citenamefont {Ruso} \emph
  {et~al.}}]{Ruso:2022qes}%
  \BibitemOpen
  \bibfield  {author} {\bibinfo {author} {\bibfnamefont {L.~A.}\ \bibnamefont
  {Ruso}} \emph {et~al.},\ }\href {\doibase 10.1088/1361-6471/adae26}
  {\bibfield  {journal} {\bibinfo  {journal} {J. Phys. G}\ }\textbf {\bibinfo
  {volume} {52}},\ \bibinfo {pages} {043001} (\bibinfo {year} {2025})},\
  \Eprint {http://arxiv.org/abs/2203.09030} {arXiv:2203.09030} \BibitemShut
  {NoStop}%
\bibitem [{\citenamefont {Hill}\ and\ \citenamefont
  {Plestid}(2024{\natexlab{a}})}]{Hill:2023acw}%
  \BibitemOpen
  \bibfield  {author} {\bibinfo {author} {\bibfnamefont {R.~J.}\ \bibnamefont
  {Hill}}\ and\ \bibinfo {author} {\bibfnamefont {R.}~\bibnamefont {Plestid}},\
  }\href {\doibase 10.1103/PhysRevLett.133.021803} {\bibfield  {journal}
  {\bibinfo  {journal} {Phys. Rev. Lett.}\ }\textbf {\bibinfo {volume} {133}},\
  \bibinfo {pages} {021803} (\bibinfo {year} {2024}{\natexlab{a}})},\ \Eprint
  {http://arxiv.org/abs/2309.07343} {arXiv:2309.07343} \BibitemShut {NoStop}%
\bibitem [{\citenamefont {Hill}\ and\ \citenamefont
  {Plestid}(2024{\natexlab{b}})}]{Hill:2023bfh}%
  \BibitemOpen
  \bibfield  {author} {\bibinfo {author} {\bibfnamefont {R.~J.}\ \bibnamefont
  {Hill}}\ and\ \bibinfo {author} {\bibfnamefont {R.}~\bibnamefont {Plestid}},\
  }\href {\doibase 10.1103/PhysRevD.109.056006} {\bibfield  {journal} {\bibinfo
   {journal} {Phys. Rev. D}\ }\textbf {\bibinfo {volume} {109}},\ \bibinfo
  {pages} {056006} (\bibinfo {year} {2024}{\natexlab{b}})},\ \Eprint
  {http://arxiv.org/abs/2309.15929} {arXiv:2309.15929} \BibitemShut {NoStop}%
\bibitem [{\citenamefont {Cirigliano}\ \emph {et~al.}(2023)\citenamefont
  {Cirigliano}, \citenamefont {Dekens}, \citenamefont {Mereghetti},\ and\
  \citenamefont {Tomalak}}]{Cirigliano:2023fnz}%
  \BibitemOpen
  \bibfield  {author} {\bibinfo {author} {\bibfnamefont {V.}~\bibnamefont
  {Cirigliano}}, \bibinfo {author} {\bibfnamefont {W.}~\bibnamefont {Dekens}},
  \bibinfo {author} {\bibfnamefont {E.}~\bibnamefont {Mereghetti}}, \ and\
  \bibinfo {author} {\bibfnamefont {O.}~\bibnamefont {Tomalak}},\ }\href
  {\doibase 10.1103/PhysRevD.108.053003} {\bibfield  {journal} {\bibinfo
  {journal} {Phys. Rev. D}\ }\textbf {\bibinfo {volume} {108}},\ \bibinfo
  {pages} {053003} (\bibinfo {year} {2023})},\ \Eprint
  {http://arxiv.org/abs/2306.03138} {arXiv:2306.03138} \BibitemShut {NoStop}%
\bibitem [{\citenamefont {Cirigliano}\ \emph {et~al.}(2025)\citenamefont
  {Cirigliano}, \citenamefont {Dekens}, \citenamefont {Mereghetti},\ and\
  \citenamefont {Tomalak}}]{Cirigliano:2024nfi}%
  \BibitemOpen
  \bibfield  {author} {\bibinfo {author} {\bibfnamefont {V.}~\bibnamefont
  {Cirigliano}}, \bibinfo {author} {\bibfnamefont {W.}~\bibnamefont {Dekens}},
  \bibinfo {author} {\bibfnamefont {E.}~\bibnamefont {Mereghetti}}, \ and\
  \bibinfo {author} {\bibfnamefont {O.}~\bibnamefont {Tomalak}},\ }\href
  {\doibase 10.1103/PhysRevD.111.053005} {\bibfield  {journal} {\bibinfo
  {journal} {Phys. Rev. D}\ }\textbf {\bibinfo {volume} {111}},\ \bibinfo
  {pages} {053005} (\bibinfo {year} {2025})},\ \Eprint
  {http://arxiv.org/abs/2410.21404} {arXiv:2410.21404} \BibitemShut {NoStop}%
\bibitem [{\citenamefont {Cirigliano}\ \emph {et~al.}(2024)\citenamefont
  {Cirigliano}, \citenamefont {Dekens}, \citenamefont {de~Vries}, \citenamefont
  {Gandolfi}, \citenamefont {Hoferichter},\ and\ \citenamefont
  {Mereghetti}}]{Cirigliano:2024rfk}%
  \BibitemOpen
  \bibfield  {author} {\bibinfo {author} {\bibfnamefont {V.}~\bibnamefont
  {Cirigliano}}, \bibinfo {author} {\bibfnamefont {W.}~\bibnamefont {Dekens}},
  \bibinfo {author} {\bibfnamefont {J.}~\bibnamefont {de~Vries}}, \bibinfo
  {author} {\bibfnamefont {S.}~\bibnamefont {Gandolfi}}, \bibinfo {author}
  {\bibfnamefont {M.}~\bibnamefont {Hoferichter}}, \ and\ \bibinfo {author}
  {\bibfnamefont {E.}~\bibnamefont {Mereghetti}},\ }\href {\doibase
  10.1103/PhysRevLett.133.211801} {\bibfield  {journal} {\bibinfo  {journal}
  {Phys. Rev. Lett.}\ }\textbf {\bibinfo {volume} {133}},\ \bibinfo {pages}
  {211801} (\bibinfo {year} {2024})},\ \Eprint
  {http://arxiv.org/abs/2405.18469} {arXiv:2405.18469} \BibitemShut {NoStop}%
\bibitem [{\citenamefont {Borah}\ \emph {et~al.}(2024)\citenamefont {Borah},
  \citenamefont {Hill},\ and\ \citenamefont {Plestid}}]{Borah:2024ghn}%
  \BibitemOpen
  \bibfield  {author} {\bibinfo {author} {\bibfnamefont {K.}~\bibnamefont
  {Borah}}, \bibinfo {author} {\bibfnamefont {R.~J.}\ \bibnamefont {Hill}}, \
  and\ \bibinfo {author} {\bibfnamefont {R.}~\bibnamefont {Plestid}},\ }\href
  {\doibase 10.1103/PhysRevD.109.113007} {\bibfield  {journal} {\bibinfo
  {journal} {Phys. Rev. D}\ }\textbf {\bibinfo {volume} {109}},\ \bibinfo
  {pages} {113007} (\bibinfo {year} {2024})},\ \Eprint
  {http://arxiv.org/abs/2402.13307} {arXiv:2402.13307} \BibitemShut {NoStop}%
\bibitem [{\citenamefont {Vander~Griend}\ \emph {et~al.}(2025)\citenamefont
  {Vander~Griend}, \citenamefont {Cao}, \citenamefont {Hill},\ and\
  \citenamefont {Plestid}}]{VanderGriend:2025mdc}%
  \BibitemOpen
  \bibfield  {author} {\bibinfo {author} {\bibfnamefont {P.}~\bibnamefont
  {Vander~Griend}}, \bibinfo {author} {\bibfnamefont {Z.}~\bibnamefont {Cao}},
  \bibinfo {author} {\bibfnamefont {R.}~\bibnamefont {Hill}}, \ and\ \bibinfo
  {author} {\bibfnamefont {R.}~\bibnamefont {Plestid}},\ }\href@noop {} {\
  (\bibinfo {year} {2025})},\ \Eprint {http://arxiv.org/abs/2501.17916}
  {arXiv:2501.17916} \BibitemShut {NoStop}%
\bibitem [{\citenamefont {Afanasev}\ \emph {et~al.}(2024)\citenamefont
  {Afanasev} \emph {et~al.}}]{Afanasev:2023gev}%
  \BibitemOpen
  \bibfield  {author} {\bibinfo {author} {\bibfnamefont {A.}~\bibnamefont
  {Afanasev}} \emph {et~al.},\ }\href {\doibase
  10.1140/epja/s10050-024-01281-y} {\bibfield  {journal} {\bibinfo  {journal}
  {Eur. Phys. J. A}\ }\textbf {\bibinfo {volume} {60}},\ \bibinfo {pages} {91}
  (\bibinfo {year} {2024})},\ \Eprint {http://arxiv.org/abs/2306.14578}
  {arXiv:2306.14578} \BibitemShut {NoStop}%
\bibitem [{\citenamefont {Richardson}\ and\ \citenamefont
  {Reis}(2024)}]{Richardson:2023vyf}%
  \BibitemOpen
  \bibfield  {author} {\bibinfo {author} {\bibfnamefont {T.~R.}\ \bibnamefont
  {Richardson}}\ and\ \bibinfo {author} {\bibfnamefont {I.~C.}\ \bibnamefont
  {Reis}},\ }\href {\doibase 10.1007/s00601-024-01948-8} {\bibfield  {journal}
  {\bibinfo  {journal} {Few Body Syst.}\ }\textbf {\bibinfo {volume} {65}},\
  \bibinfo {pages} {79} (\bibinfo {year} {2024})},\ \Eprint
  {http://arxiv.org/abs/2309.16385} {arXiv:2309.16385} \BibitemShut {NoStop}%
\bibitem [{\citenamefont {Seng}\ \emph {et~al.}(2025)\citenamefont {Seng},
  \citenamefont {Glick-Magid},\ and\ \citenamefont
  {Cirigliano}}]{Seng:2024zuc}%
  \BibitemOpen
  \bibfield  {author} {\bibinfo {author} {\bibfnamefont {C.-Y.}\ \bibnamefont
  {Seng}}, \bibinfo {author} {\bibfnamefont {A.}~\bibnamefont {Glick-Magid}}, \
  and\ \bibinfo {author} {\bibfnamefont {V.}~\bibnamefont {Cirigliano}},\
  }\href {\doibase 10.1103/PhysRevLett.134.081805} {\bibfield  {journal}
  {\bibinfo  {journal} {Phys. Rev. Lett.}\ }\textbf {\bibinfo {volume} {134}},\
  \bibinfo {pages} {081805} (\bibinfo {year} {2025})},\ \Eprint
  {http://arxiv.org/abs/2409.18115} {arXiv:2409.18115} \BibitemShut {NoStop}%
\bibitem [{\citenamefont {Combes}\ \emph {et~al.}(2024)\citenamefont {Combes},
  \citenamefont {Mereghetti},\ and\ \citenamefont {Platter}}]{Combes:2024pvm}%
  \BibitemOpen
  \bibfield  {author} {\bibinfo {author} {\bibfnamefont {E.}~\bibnamefont
  {Combes}}, \bibinfo {author} {\bibfnamefont {E.}~\bibnamefont {Mereghetti}},
  \ and\ \bibinfo {author} {\bibfnamefont {L.}~\bibnamefont {Platter}},\ }\href
  {\doibase 10.1103/PhysRevC.110.L041001} {\bibfield  {journal} {\bibinfo
  {journal} {Phys. Rev. C}\ }\textbf {\bibinfo {volume} {110}},\ \bibinfo
  {pages} {L041001} (\bibinfo {year} {2024})},\ \Eprint
  {http://arxiv.org/abs/2407.08015} {arXiv:2407.08015} \BibitemShut {NoStop}%
\bibitem [{\citenamefont {Bottino}\ \emph {et~al.}(1966)\citenamefont
  {Bottino}, \citenamefont {Ciocchetti},\ and\ \citenamefont
  {Molinari}}]{BOTTINO1966192}%
  \BibitemOpen
  \bibfield  {author} {\bibinfo {author} {\bibfnamefont {A.}~\bibnamefont
  {Bottino}}, \bibinfo {author} {\bibfnamefont {G.}~\bibnamefont {Ciocchetti}},
  \ and\ \bibinfo {author} {\bibfnamefont {A.}~\bibnamefont {Molinari}},\
  }\href {\doibase https://doi.org/10.1016/0029-5582(66)90854-6} {\bibfield
  {journal} {\bibinfo  {journal} {Nuclear Physics}\ }\textbf {\bibinfo {volume}
  {89}},\ \bibinfo {pages} {192} (\bibinfo {year} {1966})}\BibitemShut
  {NoStop}%
\bibitem [{\citenamefont {Bethe}\ and\ \citenamefont
  {Molinari}(1971)}]{Bethe:1971es}%
  \BibitemOpen
  \bibfield  {author} {\bibinfo {author} {\bibfnamefont {H.~A.}\ \bibnamefont
  {Bethe}}\ and\ \bibinfo {author} {\bibfnamefont {A.}~\bibnamefont
  {Molinari}},\ }\href {\doibase 10.1016/0003-4916(71)90019-4} {\bibfield
  {journal} {\bibinfo  {journal} {Annals Phys.}\ }\textbf {\bibinfo {volume}
  {63}},\ \bibinfo {pages} {393} (\bibinfo {year} {1971})}\BibitemShut
  {NoStop}%
\bibitem [{\citenamefont {Lin}(1972)}]{Lin:1972ba}%
  \BibitemOpen
  \bibfield  {author} {\bibinfo {author} {\bibfnamefont {W.~F.}\ \bibnamefont
  {Lin}},\ }\href {\doibase 10.1016/0370-2693(72)90314-0} {\bibfield  {journal}
  {\bibinfo  {journal} {Phys. Lett. B}\ }\textbf {\bibinfo {volume} {39}},\
  \bibinfo {pages} {447} (\bibinfo {year} {1972})}\BibitemShut {NoStop}%
\bibitem [{\citenamefont {Friar}\ and\ \citenamefont
  {Rosen}(1974)}]{Friar:1974bn}%
  \BibitemOpen
  \bibfield  {author} {\bibinfo {author} {\bibfnamefont {J.~L.}\ \bibnamefont
  {Friar}}\ and\ \bibinfo {author} {\bibfnamefont {M.}~\bibnamefont {Rosen}},\
  }\href {\doibase 10.1016/0003-4916(74)90038-4} {\bibfield  {journal}
  {\bibinfo  {journal} {Annals Phys.}\ }\textbf {\bibinfo {volume} {87}},\
  \bibinfo {pages} {289} (\bibinfo {year} {1974})}\BibitemShut {NoStop}%
\bibitem [{\citenamefont {Offermann}\ \emph {et~al.}(1991)\citenamefont
  {Offermann}, \citenamefont {Cardman}, \citenamefont {de~Jager}, \citenamefont
  {Miska}, \citenamefont {de~Vries},\ and\ \citenamefont
  {de~Vries}}]{Offermann:1991ft}%
  \BibitemOpen
  \bibfield  {author} {\bibinfo {author} {\bibfnamefont {E.~A. J.~M.}\
  \bibnamefont {Offermann}}, \bibinfo {author} {\bibfnamefont {L.~S.}\
  \bibnamefont {Cardman}}, \bibinfo {author} {\bibfnamefont {C.~W.}\
  \bibnamefont {de~Jager}}, \bibinfo {author} {\bibfnamefont {H.}~\bibnamefont
  {Miska}}, \bibinfo {author} {\bibfnamefont {C.}~\bibnamefont {de~Vries}}, \
  and\ \bibinfo {author} {\bibfnamefont {H.}~\bibnamefont {de~Vries}},\ }\href
  {\doibase 10.1103/PhysRevC.44.1096} {\bibfield  {journal} {\bibinfo
  {journal} {Phys. Rev. C}\ }\textbf {\bibinfo {volume} {44}},\ \bibinfo
  {pages} {1096} (\bibinfo {year} {1991})}\BibitemShut {NoStop}%
\bibitem [{\citenamefont {Afanasev}\ \emph {et~al.}(2005)\citenamefont
  {Afanasev}, \citenamefont {Brodsky}, \citenamefont {Carlson}, \citenamefont
  {Chen},\ and\ \citenamefont {Vanderhaeghen}}]{Afanasev:2005mp}%
  \BibitemOpen
  \bibfield  {author} {\bibinfo {author} {\bibfnamefont {A.~V.}\ \bibnamefont
  {Afanasev}}, \bibinfo {author} {\bibfnamefont {S.~J.}\ \bibnamefont
  {Brodsky}}, \bibinfo {author} {\bibfnamefont {C.~E.}\ \bibnamefont
  {Carlson}}, \bibinfo {author} {\bibfnamefont {Y.-C.}\ \bibnamefont {Chen}}, \
  and\ \bibinfo {author} {\bibfnamefont {M.}~\bibnamefont {Vanderhaeghen}},\
  }\href {\doibase 10.1103/PhysRevD.72.013008} {\bibfield  {journal} {\bibinfo
  {journal} {Phys. Rev. D}\ }\textbf {\bibinfo {volume} {72}},\ \bibinfo
  {pages} {013008} (\bibinfo {year} {2005})},\ \Eprint
  {http://arxiv.org/abs/hep-ph/0502013} {arXiv:hep-ph/0502013} \BibitemShut
  {NoStop}%
\bibitem [{\citenamefont {Gorchtein}\ \emph {et~al.}(2011)\citenamefont
  {Gorchtein}, \citenamefont {Horowitz},\ and\ \citenamefont
  {Ramsey-Musolf}}]{Gorchtein:2011mz}%
  \BibitemOpen
  \bibfield  {author} {\bibinfo {author} {\bibfnamefont {M.}~\bibnamefont
  {Gorchtein}}, \bibinfo {author} {\bibfnamefont {C.~J.}\ \bibnamefont
  {Horowitz}}, \ and\ \bibinfo {author} {\bibfnamefont {M.~J.}\ \bibnamefont
  {Ramsey-Musolf}},\ }\href {\doibase 10.1103/PhysRevC.84.015502} {\bibfield
  {journal} {\bibinfo  {journal} {Phys. Rev. C}\ }\textbf {\bibinfo {volume}
  {84}},\ \bibinfo {pages} {015502} (\bibinfo {year} {2011})},\ \Eprint
  {http://arxiv.org/abs/1102.3910} {arXiv:1102.3910} \BibitemShut {NoStop}%
\bibitem [{\citenamefont {Arrington}\ \emph {et~al.}(2011)\citenamefont
  {Arrington}, \citenamefont {Blunden},\ and\ \citenamefont
  {Melnitchouk}}]{Arrington:2011dn}%
  \BibitemOpen
  \bibfield  {author} {\bibinfo {author} {\bibfnamefont {J.}~\bibnamefont
  {Arrington}}, \bibinfo {author} {\bibfnamefont {P.~G.}\ \bibnamefont
  {Blunden}}, \ and\ \bibinfo {author} {\bibfnamefont {W.}~\bibnamefont
  {Melnitchouk}},\ }\href {\doibase 10.1016/j.ppnp.2011.07.003} {\bibfield
  {journal} {\bibinfo  {journal} {Prog. Part. Nucl. Phys.}\ }\textbf {\bibinfo
  {volume} {66}},\ \bibinfo {pages} {782} (\bibinfo {year} {2011})},\ \Eprint
  {http://arxiv.org/abs/1105.0951} {arXiv:1105.0951} \BibitemShut {NoStop}%
\bibitem [{\citenamefont {Seng}\ \emph {et~al.}(2019)\citenamefont {Seng},
  \citenamefont {Gorchtein},\ and\ \citenamefont
  {Ramsey-Musolf}}]{Seng:2018qru}%
  \BibitemOpen
  \bibfield  {author} {\bibinfo {author} {\bibfnamefont {C.~Y.}\ \bibnamefont
  {Seng}}, \bibinfo {author} {\bibfnamefont {M.}~\bibnamefont {Gorchtein}}, \
  and\ \bibinfo {author} {\bibfnamefont {M.~J.}\ \bibnamefont
  {Ramsey-Musolf}},\ }\href {\doibase 10.1103/PhysRevD.100.013001} {\bibfield
  {journal} {\bibinfo  {journal} {Phys. Rev. D}\ }\textbf {\bibinfo {volume}
  {100}},\ \bibinfo {pages} {013001} (\bibinfo {year} {2019})},\ \Eprint
  {http://arxiv.org/abs/1812.03352} {arXiv:1812.03352} \BibitemShut {NoStop}%
\bibitem [{\citenamefont {Pasquini}\ and\ \citenamefont
  {Vanderhaeghen}(2018)}]{Pasquini:2018wbl}%
  \BibitemOpen
  \bibfield  {author} {\bibinfo {author} {\bibfnamefont {B.}~\bibnamefont
  {Pasquini}}\ and\ \bibinfo {author} {\bibfnamefont {M.}~\bibnamefont
  {Vanderhaeghen}},\ }\href {\doibase 10.1146/annurev-nucl-101917-020843}
  {\bibfield  {journal} {\bibinfo  {journal} {Ann. Rev. Nucl. Part. Sci.}\
  }\textbf {\bibinfo {volume} {68}},\ \bibinfo {pages} {75} (\bibinfo {year}
  {2018})},\ \Eprint {http://arxiv.org/abs/1805.10482} {arXiv:1805.10482}
  \BibitemShut {NoStop}%
\bibitem [{\citenamefont {Gu\`eye}\ \emph {et~al.}(2020)\citenamefont {Gu\`eye}
  \emph {et~al.}}]{JeffersonLabHallA:2018cmf}%
  \BibitemOpen
  \bibfield  {author} {\bibinfo {author} {\bibfnamefont {P.}~\bibnamefont
  {Gu\`eye}} \emph {et~al.} (\bibinfo {collaboration} {Jefferson Lab Hall A}),\
  }\href {\doibase 10.1140/epja/s10050-020-00135-7} {\bibfield  {journal}
  {\bibinfo  {journal} {Eur. Phys. J. A}\ }\textbf {\bibinfo {volume} {56}},\
  \bibinfo {pages} {126} (\bibinfo {year} {2020})},\ \bibinfo {note} {[Erratum:
  Eur.Phys.J.A 56, 228 (2020)]},\ \Eprint {http://arxiv.org/abs/1805.12441}
  {arXiv:1805.12441} \BibitemShut {NoStop}%
\bibitem [{\citenamefont {Kutz}\ and\ \citenamefont
  {Schmidt}(2022)}]{Kutz:2022hka}%
  \BibitemOpen
  \bibfield  {author} {\bibinfo {author} {\bibfnamefont {T.}~\bibnamefont
  {Kutz}}\ and\ \bibinfo {author} {\bibfnamefont {A.}~\bibnamefont {Schmidt}},\
  }\href {\doibase 10.1140/epja/s10050-022-00682-1} {\bibfield  {journal}
  {\bibinfo  {journal} {Eur. Phys. J. A}\ }\textbf {\bibinfo {volume} {58}},\
  \bibinfo {pages} {36} (\bibinfo {year} {2022})}\BibitemShut {NoStop}%
\bibitem [{\citenamefont {Jakubassa-Amundsen}\ and\ \citenamefont
  {Roca-Maza}(2023)}]{Jakubassa-Amundsen:2023wuz}%
  \BibitemOpen
  \bibfield  {author} {\bibinfo {author} {\bibfnamefont {D.~H.}\ \bibnamefont
  {Jakubassa-Amundsen}}\ and\ \bibinfo {author} {\bibfnamefont
  {X.}~\bibnamefont {Roca-Maza}},\ }\href {\doibase
  10.1103/PhysRevC.108.034314} {\bibfield  {journal} {\bibinfo  {journal}
  {Phys. Rev. C}\ }\textbf {\bibinfo {volume} {108}},\ \bibinfo {pages}
  {034314} (\bibinfo {year} {2023})},\ \Eprint
  {http://arxiv.org/abs/2307.13469} {arXiv:2307.13469} \BibitemShut {NoStop}%
\bibitem [{\citenamefont
  {Jakubassa-Amundsen}(2024)}]{Jakubassa-Amundsen:2024dai}%
  \BibitemOpen
  \bibfield  {author} {\bibinfo {author} {\bibfnamefont {D.~H.}\ \bibnamefont
  {Jakubassa-Amundsen}},\ }\href {\doibase 10.1103/PhysRevC.109.L061303}
  {\bibfield  {journal} {\bibinfo  {journal} {Phys. Rev. C}\ }\textbf {\bibinfo
  {volume} {109}},\ \bibinfo {pages} {L061303} (\bibinfo {year} {2024})},\
  \bibinfo {note} {[Erratum: Phys.Rev.C 110, 069902 (2024)]}\BibitemShut
  {NoStop}%
\bibitem [{\citenamefont {Gennari}\ \emph {et~al.}(2025)\citenamefont
  {Gennari}, \citenamefont {Drissi}, \citenamefont {Gorchtein}, \citenamefont
  {Navratil},\ and\ \citenamefont {Seng}}]{Gennari:2024sbn}%
  \BibitemOpen
  \bibfield  {author} {\bibinfo {author} {\bibfnamefont {M.}~\bibnamefont
  {Gennari}}, \bibinfo {author} {\bibfnamefont {M.}~\bibnamefont {Drissi}},
  \bibinfo {author} {\bibfnamefont {M.}~\bibnamefont {Gorchtein}}, \bibinfo
  {author} {\bibfnamefont {P.}~\bibnamefont {Navratil}}, \ and\ \bibinfo
  {author} {\bibfnamefont {C.-Y.}\ \bibnamefont {Seng}},\ }\href {\doibase
  10.1103/PhysRevLett.134.012501} {\bibfield  {journal} {\bibinfo  {journal}
  {Phys. Rev. Lett.}\ }\textbf {\bibinfo {volume} {134}},\ \bibinfo {pages}
  {012501} (\bibinfo {year} {2025})},\ \Eprint
  {http://arxiv.org/abs/2405.19281} {arXiv:2405.19281} \BibitemShut {NoStop}%
\bibitem [{\citenamefont {Siegert}(1937)}]{PhysRev.52.787}%
  \BibitemOpen
  \bibfield  {author} {\bibinfo {author} {\bibfnamefont {A.~J.~F.}\
  \bibnamefont {Siegert}},\ }\href {\doibase 10.1103/PhysRev.52.787} {\bibfield
   {journal} {\bibinfo  {journal} {Phys. Rev.}\ }\textbf {\bibinfo {volume}
  {52}},\ \bibinfo {pages} {787} (\bibinfo {year} {1937})}\BibitemShut
  {NoStop}%
\bibitem [{\citenamefont
  {Jakubassa-Amundsen}(2022)}]{Jakubassa-Amundsen:2022ckf}%
  \BibitemOpen
  \bibfield  {author} {\bibinfo {author} {\bibfnamefont {D.~H.}\ \bibnamefont
  {Jakubassa-Amundsen}},\ }\href {\doibase 10.1103/PhysRevC.105.054303}
  {\bibfield  {journal} {\bibinfo  {journal} {Phys. Rev. C}\ }\textbf {\bibinfo
  {volume} {105}},\ \bibinfo {pages} {054303} (\bibinfo {year}
  {2022})}\BibitemShut {NoStop}%
\bibitem [{\citenamefont {Pastore}\ \emph {et~al.}(2008)\citenamefont
  {Pastore}, \citenamefont {Schiavilla},\ and\ \citenamefont
  {Goity}}]{Pastore:2008ui}%
  \BibitemOpen
  \bibfield  {author} {\bibinfo {author} {\bibfnamefont {S.}~\bibnamefont
  {Pastore}}, \bibinfo {author} {\bibfnamefont {R.}~\bibnamefont {Schiavilla}},
  \ and\ \bibinfo {author} {\bibfnamefont {J.~L.}\ \bibnamefont {Goity}},\
  }\href {\doibase 10.1103/PhysRevC.78.064002} {\bibfield  {journal} {\bibinfo
  {journal} {Phys. Rev. C}\ }\textbf {\bibinfo {volume} {78}},\ \bibinfo
  {pages} {064002} (\bibinfo {year} {2008})},\ \Eprint
  {http://arxiv.org/abs/0810.1941} {arXiv:0810.1941} \BibitemShut {NoStop}%
\bibitem [{\citenamefont {Pastore}\ \emph {et~al.}(2009)\citenamefont
  {Pastore}, \citenamefont {Girlanda}, \citenamefont {Schiavilla},
  \citenamefont {Viviani},\ and\ \citenamefont {Wiringa}}]{Pastore:2009is}%
  \BibitemOpen
  \bibfield  {author} {\bibinfo {author} {\bibfnamefont {S.}~\bibnamefont
  {Pastore}}, \bibinfo {author} {\bibfnamefont {L.}~\bibnamefont {Girlanda}},
  \bibinfo {author} {\bibfnamefont {R.}~\bibnamefont {Schiavilla}}, \bibinfo
  {author} {\bibfnamefont {M.}~\bibnamefont {Viviani}}, \ and\ \bibinfo
  {author} {\bibfnamefont {R.~B.}\ \bibnamefont {Wiringa}},\ }\href {\doibase
  10.1103/PhysRevC.80.034004} {\bibfield  {journal} {\bibinfo  {journal} {Phys.
  Rev. C}\ }\textbf {\bibinfo {volume} {80}},\ \bibinfo {pages} {034004}
  (\bibinfo {year} {2009})},\ \Eprint {http://arxiv.org/abs/0906.1800}
  {arXiv:0906.1800} \BibitemShut {NoStop}%
\bibitem [{\citenamefont {Pastore}\ \emph {et~al.}(2011)\citenamefont
  {Pastore}, \citenamefont {Girlanda}, \citenamefont {Schiavilla},\ and\
  \citenamefont {Viviani}}]{Pastore:2011ip}%
  \BibitemOpen
  \bibfield  {author} {\bibinfo {author} {\bibfnamefont {S.}~\bibnamefont
  {Pastore}}, \bibinfo {author} {\bibfnamefont {L.}~\bibnamefont {Girlanda}},
  \bibinfo {author} {\bibfnamefont {R.}~\bibnamefont {Schiavilla}}, \ and\
  \bibinfo {author} {\bibfnamefont {M.}~\bibnamefont {Viviani}},\ }\href
  {\doibase 10.1103/PhysRevC.84.024001} {\bibfield  {journal} {\bibinfo
  {journal} {Phys. Rev. C}\ }\textbf {\bibinfo {volume} {84}},\ \bibinfo
  {pages} {024001} (\bibinfo {year} {2011})},\ \Eprint
  {http://arxiv.org/abs/1106.4539} {arXiv:1106.4539} \BibitemShut {NoStop}%
\bibitem [{\citenamefont {Kolling}\ \emph {et~al.}(2011)\citenamefont
  {Kolling}, \citenamefont {Epelbaum}, \citenamefont {Krebs},\ and\
  \citenamefont {Meissner}}]{Kolling:2011mt}%
  \BibitemOpen
  \bibfield  {author} {\bibinfo {author} {\bibfnamefont {S.}~\bibnamefont
  {Kolling}}, \bibinfo {author} {\bibfnamefont {E.}~\bibnamefont {Epelbaum}},
  \bibinfo {author} {\bibfnamefont {H.}~\bibnamefont {Krebs}}, \ and\ \bibinfo
  {author} {\bibfnamefont {U.~G.}\ \bibnamefont {Meissner}},\ }\href {\doibase
  10.1103/PhysRevC.84.054008} {\bibfield  {journal} {\bibinfo  {journal} {Phys.
  Rev. C}\ }\textbf {\bibinfo {volume} {84}},\ \bibinfo {pages} {054008}
  (\bibinfo {year} {2011})},\ \Eprint {http://arxiv.org/abs/1107.0602}
  {arXiv:1107.0602} \BibitemShut {NoStop}%
\bibitem [{\citenamefont {Baroni}\ \emph {et~al.}(2021)\citenamefont {Baroni},
  \citenamefont {King},\ and\ \citenamefont {Pastore}}]{Baroni:2021vrc}%
  \BibitemOpen
  \bibfield  {author} {\bibinfo {author} {\bibfnamefont {A.}~\bibnamefont
  {Baroni}}, \bibinfo {author} {\bibfnamefont {G.~B.}\ \bibnamefont {King}}, \
  and\ \bibinfo {author} {\bibfnamefont {S.}~\bibnamefont {Pastore}},\ }\href
  {\doibase 10.1007/s00601-021-01700-6} {\bibfield  {journal} {\bibinfo
  {journal} {Few Body Syst.}\ }\textbf {\bibinfo {volume} {62}},\ \bibinfo
  {pages} {114} (\bibinfo {year} {2021})},\ \Eprint
  {http://arxiv.org/abs/2107.10721} {arXiv:2107.10721} \BibitemShut {NoStop}%
\bibitem [{\citenamefont {Schwinger}(1959)}]{Schwinger:1959xd}%
  \BibitemOpen
  \bibfield  {author} {\bibinfo {author} {\bibfnamefont {J.~S.}\ \bibnamefont
  {Schwinger}},\ }\href {\doibase 10.1103/PhysRevLett.3.296} {\bibfield
  {journal} {\bibinfo  {journal} {Phys. Rev. Lett.}\ }\textbf {\bibinfo
  {volume} {3}},\ \bibinfo {pages} {296} (\bibinfo {year} {1959})}\BibitemShut
  {NoStop}%
\bibitem [{\citenamefont {Boulware}\ and\ \citenamefont
  {Deser}(1966)}]{Boulware:1966qsd}%
  \BibitemOpen
  \bibfield  {author} {\bibinfo {author} {\bibfnamefont {D.~G.}\ \bibnamefont
  {Boulware}}\ and\ \bibinfo {author} {\bibfnamefont {S.}~\bibnamefont
  {Deser}},\ }\href {\doibase 10.1103/PhysRev.151.1278} {\bibfield  {journal}
  {\bibinfo  {journal} {Phys. Rev.}\ }\textbf {\bibinfo {volume} {151}},\
  \bibinfo {pages} {1278} (\bibinfo {year} {1966})}\BibitemShut {NoStop}%
\bibitem [{\citenamefont {Brown}(1966)}]{Brown:1966zza}%
  \BibitemOpen
  \bibfield  {author} {\bibinfo {author} {\bibfnamefont {L.~S.}\ \bibnamefont
  {Brown}},\ }\href {\doibase 10.1103/PhysRev.150.1338} {\bibfield  {journal}
  {\bibinfo  {journal} {Phys. Rev.}\ }\textbf {\bibinfo {volume} {150}},\
  \bibinfo {pages} {1338} (\bibinfo {year} {1966})}\BibitemShut {NoStop}%
\bibitem [{\citenamefont {Tarrach}(1975)}]{Tarrach:1975tu}%
  \BibitemOpen
  \bibfield  {author} {\bibinfo {author} {\bibfnamefont {R.}~\bibnamefont
  {Tarrach}},\ }\href {\doibase 10.1007/BF02894857} {\bibfield  {journal}
  {\bibinfo  {journal} {Nuovo Cim. A}\ }\textbf {\bibinfo {volume} {28}},\
  \bibinfo {pages} {409} (\bibinfo {year} {1975})}\BibitemShut {NoStop}%
\bibitem [{\citenamefont {Hutt}\ \emph {et~al.}(2000)\citenamefont {Hutt},
  \citenamefont {L'vov}, \citenamefont {Milstein},\ and\ \citenamefont
  {Schumacher}}]{Hutt:1999pz}%
  \BibitemOpen
  \bibfield  {author} {\bibinfo {author} {\bibfnamefont {M.~T.}\ \bibnamefont
  {Hutt}}, \bibinfo {author} {\bibfnamefont {A.~I.}\ \bibnamefont {L'vov}},
  \bibinfo {author} {\bibfnamefont {A.~I.}\ \bibnamefont {Milstein}}, \ and\
  \bibinfo {author} {\bibfnamefont {M.}~\bibnamefont {Schumacher}},\ }\href
  {\doibase 10.1016/S0370-1573(99)00041-1} {\bibfield  {journal} {\bibinfo
  {journal} {Phys. Rept.}\ }\textbf {\bibinfo {volume} {323}},\ \bibinfo
  {pages} {457} (\bibinfo {year} {2000})},\ \Eprint
  {http://arxiv.org/abs/nucl-th/9905026} {arXiv:nucl-th/9905026} \BibitemShut
  {NoStop}%
\bibitem [{\citenamefont {Gorchtein}(2006)}]{Gorchtein:2005za}%
  \BibitemOpen
  \bibfield  {author} {\bibinfo {author} {\bibfnamefont {M.}~\bibnamefont
  {Gorchtein}},\ }\href {\doibase 10.1103/PhysRevC.73.035213} {\bibfield
  {journal} {\bibinfo  {journal} {Phys. Rev. C}\ }\textbf {\bibinfo {volume}
  {73}},\ \bibinfo {pages} {035213} (\bibinfo {year} {2006})},\ \Eprint
  {http://arxiv.org/abs/hep-ph/0512106} {arXiv:hep-ph/0512106} \BibitemShut
  {NoStop}%
\bibitem [{\citenamefont {Hill}\ and\ \citenamefont
  {Paz}(2017)}]{Hill:2016bjv}%
  \BibitemOpen
  \bibfield  {author} {\bibinfo {author} {\bibfnamefont {R.~J.}\ \bibnamefont
  {Hill}}\ and\ \bibinfo {author} {\bibfnamefont {G.}~\bibnamefont {Paz}},\
  }\href {\doibase 10.1103/PhysRevD.95.094017} {\bibfield  {journal} {\bibinfo
  {journal} {Phys. Rev. D}\ }\textbf {\bibinfo {volume} {95}},\ \bibinfo
  {pages} {094017} (\bibinfo {year} {2017})},\ \Eprint
  {http://arxiv.org/abs/1611.09917} {arXiv:1611.09917} \BibitemShut {NoStop}%
\bibitem [{\citenamefont {Hardy}\ and\ \citenamefont
  {Towner}(2020)}]{Hardy:2020qwl}%
  \BibitemOpen
  \bibfield  {author} {\bibinfo {author} {\bibfnamefont {J.~C.}\ \bibnamefont
  {Hardy}}\ and\ \bibinfo {author} {\bibfnamefont {I.~S.}\ \bibnamefont
  {Towner}},\ }\href {\doibase 10.1103/PhysRevC.102.045501} {\bibfield
  {journal} {\bibinfo  {journal} {Phys. Rev. C}\ }\textbf {\bibinfo {volume}
  {102}},\ \bibinfo {pages} {045501} (\bibinfo {year} {2020})}\BibitemShut
  {NoStop}%
\bibitem [{\citenamefont {Adhikari}\ \emph {et~al.}(2022)\citenamefont
  {Adhikari} \emph {et~al.}}]{CREX:2022kgg}%
  \BibitemOpen
  \bibfield  {author} {\bibinfo {author} {\bibfnamefont {D.}~\bibnamefont
  {Adhikari}} \emph {et~al.} (\bibinfo {collaboration} {CREX}),\ }\href
  {\doibase 10.1103/PhysRevLett.129.042501} {\bibfield  {journal} {\bibinfo
  {journal} {Phys. Rev. Lett.}\ }\textbf {\bibinfo {volume} {129}},\ \bibinfo
  {pages} {042501} (\bibinfo {year} {2022})},\ \Eprint
  {http://arxiv.org/abs/2205.11593} {arXiv:2205.11593} \BibitemShut {NoStop}%
\bibitem [{\citenamefont {Adhikari}\ \emph {et~al.}(2021)\citenamefont
  {Adhikari} \emph {et~al.}}]{PREX:2021umo}%
  \BibitemOpen
  \bibfield  {author} {\bibinfo {author} {\bibfnamefont {D.}~\bibnamefont
  {Adhikari}} \emph {et~al.} (\bibinfo {collaboration} {PREX}),\ }\href
  {\doibase 10.1103/PhysRevLett.126.172502} {\bibfield  {journal} {\bibinfo
  {journal} {Phys. Rev. Lett.}\ }\textbf {\bibinfo {volume} {126}},\ \bibinfo
  {pages} {172502} (\bibinfo {year} {2021})},\ \Eprint
  {http://arxiv.org/abs/2102.10767} {arXiv:2102.10767} \BibitemShut {NoStop}%
\bibitem [{\citenamefont {Epelbaum}\ \emph {et~al.}(2020)\citenamefont
  {Epelbaum}, \citenamefont {Krebs},\ and\ \citenamefont
  {Reinert}}]{Epelbaum:2019kcf}%
  \BibitemOpen
  \bibfield  {author} {\bibinfo {author} {\bibfnamefont {E.}~\bibnamefont
  {Epelbaum}}, \bibinfo {author} {\bibfnamefont {H.}~\bibnamefont {Krebs}}, \
  and\ \bibinfo {author} {\bibfnamefont {P.}~\bibnamefont {Reinert}},\ }\href
  {\doibase 10.3389/fphy.2020.00098} {\bibfield  {journal} {\bibinfo  {journal}
  {Front. in Phys.}\ }\textbf {\bibinfo {volume} {8}},\ \bibinfo {pages} {98}
  (\bibinfo {year} {2020})},\ \Eprint {http://arxiv.org/abs/1911.11875}
  {arXiv:1911.11875} \BibitemShut {NoStop}%
\bibitem [{\citenamefont {Friman-Gayer}\ \emph {et~al.}(2021)\citenamefont
  {Friman-Gayer} \emph {et~al.}}]{Friman-Gayer:2020vqn}%
  \BibitemOpen
  \bibfield  {author} {\bibinfo {author} {\bibfnamefont {U.}~\bibnamefont
  {Friman-Gayer}} \emph {et~al.},\ }\href {\doibase
  10.1103/PhysRevLett.126.102501} {\bibfield  {journal} {\bibinfo  {journal}
  {Phys. Rev. Lett.}\ }\textbf {\bibinfo {volume} {126}},\ \bibinfo {pages}
  {102501} (\bibinfo {year} {2021})},\ \Eprint
  {http://arxiv.org/abs/2005.07837} {arXiv:2005.07837} \BibitemShut {NoStop}%
\bibitem [{\citenamefont {Cirigliano}\ \emph {et~al.}(2018)\citenamefont
  {Cirigliano}, \citenamefont {Dekens}, \citenamefont {de~Vries}, \citenamefont
  {Graesser},\ and\ \citenamefont {Mereghetti}}]{Cirigliano:2018yza}%
  \BibitemOpen
  \bibfield  {author} {\bibinfo {author} {\bibfnamefont {V.}~\bibnamefont
  {Cirigliano}}, \bibinfo {author} {\bibfnamefont {W.}~\bibnamefont {Dekens}},
  \bibinfo {author} {\bibfnamefont {J.}~\bibnamefont {de~Vries}}, \bibinfo
  {author} {\bibfnamefont {M.~L.}\ \bibnamefont {Graesser}}, \ and\ \bibinfo
  {author} {\bibfnamefont {E.}~\bibnamefont {Mereghetti}},\ }\href {\doibase
  10.1007/JHEP12(2018)097} {\bibfield  {journal} {\bibinfo  {journal} {JHEP}\
  }\textbf {\bibinfo {volume} {12}},\ \bibinfo {pages} {097} (\bibinfo {year}
  {2018})},\ \Eprint {http://arxiv.org/abs/1806.02780} {arXiv:1806.02780}
  \BibitemShut {NoStop}%
\end{thebibliography}%

\end{document}